\documentclass[ALICE,manyauthors]{cernphprep}
\usepackage[comma,square,numbers,sort&compress]{natbib}
\usepackage{hyperref}
\usepackage{lineno}
\usepackage{xspace}
\usepackage[T1]{fontenc}
\usepackage{orcidlink}

\begin{document}
%

\newcommand{\pp}           {pp\xspace}
\newcommand{\ppbar}        {\mbox{$\mathrm {p\overline{p}}$}\xspace}
\newcommand{\XeXe}         {\mbox{Xe--Xe}\xspace}
\newcommand{\PbPb}         {\mbox{Pb--Pb}\xspace}
\newcommand{\pA}           {\mbox{pA}\xspace}
\newcommand{\pPb}          {\mbox{p--Pb}\xspace}
\newcommand{\AuAu}         {\mbox{Au--Au}\xspace}
\newcommand{\dAu}          {\mbox{d--Au}\xspace}

\newcommand{\s}            {\ensuremath{\sqrt{s}}\xspace}
\newcommand{\snn}          {\ensuremath{\sqrt{s_{\mathrm{NN}}}}\xspace}
\newcommand{\pt}           {\ensuremath{p_{\rm T}}\xspace}
\newcommand{\meanpt}       {$\langle p_{\mathrm{T}}\rangle$\xspace}
\newcommand{\ycms}         {\ensuremath{y_{\rm CMS}}\xspace}
\newcommand{\ylab}         {\ensuremath{y_{\rm lab}}\xspace}
\newcommand{\etarange}[1]  {\mbox{$\left | \eta \right |~<~#1$}}
\newcommand{\yrange}[1]    {\mbox{$\left | y \right |~<~#1$}}
\newcommand{\dndy}         {\ensuremath{\mathrm{d}N_\mathrm{ch}/\mathrm{d}y}\xspace}
\newcommand{\dndeta}       {\ensuremath{\mathrm{d}N_\mathrm{ch}/\mathrm{d}\eta}\xspace}
\newcommand{\avdndeta}     {\ensuremath{\langle\dndeta\rangle}\xspace}
\newcommand{\dNdy}         {\ensuremath{\mathrm{d}N_\mathrm{ch}/\mathrm{d}y}\xspace}
\newcommand{\Npart}        {\ensuremath{N_\mathrm{part}}\xspace}
\newcommand{\Ncoll}        {\ensuremath{N_\mathrm{coll}}\xspace}
\newcommand{\dEdx}         {\ensuremath{\textrm{d}E/\textrm{d}x}\xspace}
\newcommand{\RpPb}         {\ensuremath{R_{\rm pPb}}\xspace}

\newcommand{\nineH}        {$\sqrt{s}~=~0.9$~Te\kern-.1emV\xspace}
\newcommand{\seven}        {$\sqrt{s}~=~7$~Te\kern-.1emV\xspace}
\newcommand{\twoH}         {$\sqrt{s}~=~0.2$~Te\kern-.1emV\xspace}
\newcommand{\twosevensix}  {$\sqrt{s}~=~2.76$~Te\kern-.1emV\xspace}
\newcommand{\five}         {$\sqrt{s}~=~5.02$~Te\kern-.1emV\xspace}
\newcommand{\twosevensixnn}{$\sqrt{s_{\mathrm{NN}}}~=~2.76$~Te\kern-.1emV\xspace}
\newcommand{\fivenn}       {$\sqrt{s_{\mathrm{NN}}}~=~5.02$~Te\kern-.1emV\xspace}
\newcommand{\LT}           {L{\'e}vy-Tsallis\xspace}
\newcommand{\GeVc}         {Ge\kern-.1emV/$c$\xspace}
\newcommand{\MeVc}         {Me\kern-.1emV/$c$\xspace}
\newcommand{\TeV}          {Te\kern-.1emV\xspace}
\newcommand{\GeV}          {Ge\kern-.1emV\xspace}
\newcommand{\MeV}          {Me\kern-.1emV\xspace}
\newcommand{\GeVmass}      {Ge\kern-.2emV/$c^2$\xspace}
\newcommand{\MeVmass}      {Me\kern-.2emV/$c^2$\xspace}
\newcommand{\lumi}         {\ensuremath{\mathcal{L}}\xspace}

\newcommand{\ITS}          {\rm{ITS}\xspace}
\newcommand{\TOF}          {\rm{TOF}\xspace}
\newcommand{\ZDC}          {\rm{ZDC}\xspace}
\newcommand{\ZDCs}         {\rm{ZDCs}\xspace}
\newcommand{\ZNA}          {\rm{ZNA}\xspace}
\newcommand{\ZNC}          {\rm{ZNC}\xspace}
\newcommand{\SPD}          {\rm{SPD}\xspace}
\newcommand{\SDD}          {\rm{SDD}\xspace}
\newcommand{\SSD}          {\rm{SSD}\xspace}
\newcommand{\TPC}          {\rm{TPC}\xspace}
\newcommand{\TRD}          {\rm{TRD}\xspace}
\newcommand{\VZERO}        {\rm{V0}\xspace}
\newcommand{\VZEROA}       {\rm{V0A}\xspace}
\newcommand{\VZEROC}       {\rm{V0C}\xspace}
\newcommand{\Vdecay} 	   {\ensuremath{V^{0}}\xspace}

\newcommand{\ee}           {\ensuremath{e^{+}e^{-}}} 
\newcommand{\pip}          {\ensuremath{\pi^{+}}\xspace}
\newcommand{\pim}          {\ensuremath{\pi^{-}}\xspace}
\newcommand{\kap}          {\ensuremath{\rm{K}^{+}}\xspace}
\newcommand{\kam}          {\ensuremath{\rm{K}^{-}}\xspace}
\newcommand{\pbar}         {\ensuremath{\rm\overline{p}}\xspace}
\newcommand{\kzero}        {\ensuremath{{\rm K}^{0}_{\rm{S}}}\xspace}
\newcommand{\lmb}          {\ensuremath{\Lambda}\xspace}
\newcommand{\almb}         {\ensuremath{\overline{\Lambda}}\xspace}
\newcommand{\Om}           {\ensuremath{\Omega^-}\xspace}
\newcommand{\Mo}           {\ensuremath{\overline{\Omega}^+}\xspace}
\newcommand{\X}            {\ensuremath{\Xi^-}\xspace}
\newcommand{\Ix}           {\ensuremath{\overline{\Xi}^+}\xspace}
\newcommand{\Xis}          {\ensuremath{\Xi^{\pm}}\xspace}
\newcommand{\Oms}          {\ensuremath{\Omega^{\pm}}\xspace}
\newcommand{\degree}       {\ensuremath{^{\rm o}}\xspace}

\begin{titlepage}
\PHyear{2024}       
\PHnumber{146}      
\PHdate{27 May}     

\title{Investigating $\Lambda$ baryon production in p--Pb collisions in jets and the underlying event using angular correlations}   
\ShortTitle{Investigating $\Lambda$ production using angular correlations}   

\Collaboration{ALICE Collaboration\thanks{See Appendix~\ref{app:collab} for the list of collaboration members}}
\ShortAuthor{ALICE Collaboration} 

\begin{abstract}
First measurements of hadron(h)--$\Lambda$ azimuthal angular correlations in p--Pb collisions at \snn$= 5.02$~TeV using the ALICE detector at the Large Hadron Collider are presented. These correlations are used to separate the production of associated $\Lambda$ baryons into three different kinematic regions, namely those produced in the direction of the trigger particle (near-side), those produced in the opposite direction (away-side), and those whose production is uncorrelated with the jet axis (underlying event). The per-trigger associated $\Lambda$ yields in these regions are extracted, along with the near- and away-side azimuthal peak widths, and the results are studied as a function of associated particle \pt and event multiplicity. Comparisons with the DPMJET event generator and previous measurements of the $\phi(1020)$ meson are also made. The final results indicate that strangeness production in the highest multiplicity \pPb collisions is enhanced relative to low multiplicity collisions in both the jet-like regions and the underlying event. The production of \lmb relative to charged hadrons is also enhanced in the underlying event when compared to the jet-like regions. Additionally, the results hint that strange quark production in the away-side of the jet is modified by soft interactions with the underlying event.
\end{abstract}

\end{titlepage}

\setcounter{page}{2} 


\section{Introduction} 
\label{sec:introduction}

The production of strange hadrons is expected to be enhanced in heavy-ion collisions when compared to minimum bias proton--proton (pp) collisions due to the formation of a deconfined, partonic phase of matter known as the quark--gluon plasma (QGP)~\cite{QGPStrangeness}. This enhancement is caused by the yield equilibration of strange quarks in the QGP, made possible by gluon fusion ($\text{gg} \rightarrow \text{s}\bar{\text{s}}$) in the thermalized medium~\cite{StrangenessProduction1, StrangenessProduction2}. Strangeness enhancement in heavy-ion collisions with respect to minimum bias pp collisions has been experimentally verified via the measurement of the relative strange hadron to pion yields using multiple detectors at different energies, starting with the Super Proton Synchrotron (SPS)~\cite{SPSEnhancement} in the early 1990s, followed by the Relativistic Heavy Ion Collider (RHIC)~\cite{STAREnhancement} in the 2000s, and more recently with the Large Hadron Collider (LHC) in the 2010s~\cite{ALICEOmegaEnhancement}. With the exception of the most peripheral collisions, these measurements do not show a large dependence on either the collision centrality or the collision energy~\cite{ALICEppEnhancement}. However, the peripheral measurements hint at a smooth increase in the degree of strangeness equilibration as a function of collision centrality, which is not well understood. Theoretical models which rely heavily on a grand-canonical description of particle production are able to reproduce the enhancement of strange hadron to pion yields for only the most central heavy-ion collisions~\cite{GrandCanon1, GrandCanon2}. To describe the lower-multiplicity \PbPb results, further assumptions (e.g., canonical strangeness suppression ~\cite{StrangenessSuppression} or core-corona superposition ~\cite{CoreCorona}) are required.

Furthermore, measurements of strange hadron production in smaller collision systems from the ALICE Collaboration (pp, p--Pb) and the E910 experiment (p--Au) exhibit a similar enhancement relative to low-multiplicity pp collisions~\cite{ALICEppEnhancement, ALICEpPbEnhancement, E910pPbEnhancement}. In fact, the ratio of strange hadrons to pions observed in high multiplicity pp and p--Pb collisions is comparable with that observed in Pb--Pb collisions, with a smooth transition occurring across the different collision systems. Additionally, the relative amount of enhancement is observed to scale with the strangeness content of the hadron~\cite{ALICEOmegaEnhancement}. These observations challenge the initial expectation that QGP formation only occurs in heavy-ion collisions, and suggests that the underlying physics responsible for strangeness enhancement is not exclusive to large systems. These measurements in small systems are typically interpreted phenomenologically as the result of both initial-state collective dynamics in the form of color-glass condensate~\cite{ColorGlass} and final-state collective dynamics in the form of hydrodynamic flow~\cite{HydroFlow}. The latest Monte Carlo (MC) pp event generators, such as PYTHIA 8~\cite{Pythia1, Pythia2} and its Angantyr tune~\cite{PythiaAgantyr}, are able to describe the strangeness enhancement data from pp collisions using other phenomenological approaches, such as color reconnection~\cite{ColorReconnection} and string shoving~\cite{StringShoving}.  However, to further understand the underlying physics responsible for this enhancement, it is necessary to study the production of strange hadrons as a function of multiplicity in smaller systems in more detail.

Two-particle angular correlation functions~\cite{ALICEAngularCorrelations1, ALICEAngularCorrelations2} have been used in the past to measure strange particle production both in and out of jets~\cite{ALICE:2024aid, Lambda1, ALICE:2024zxp}. By correlating high transverse momentum (\pt) trigger hadrons (as proxies for the jet axes) with lower-\pt associated particles in both azimuthal angle ($\Delta\varphi$) and pseudorapidity ($\Delta\eta$), the resulting distributions exhibit clear peaks about $\Delta\varphi = 0$ and $\Delta\varphi = \pi$. These peaks are the result of jet fragmentation, and are generally referred to as the near-side and away-side peaks, respectively. The near-side peak region is often connected to associated particle production within a jet with little-to-no modification through scattering with out-of-jet particles, as selecting for high-\pt trigger hadrons biases the collision event sample towards those with hard-scattering processes that occur near the surface of the collision region~\cite{SurfaceBias}. The away-side region corresponds to particle production within the recoiling jet that has potentially been modified by interactions with the softer particles produced out of jet. These softer out-of-jet particles are referred to as the ``underlying event'' (UE), and they are mostly uncorrelated in ($\Delta\varphi$, $\Delta\eta$). See Section~\ref{sec:uefit} for the experimental definition of the UE utilized in this analysis. The production of these UE particles is dominated by multi-parton interactions (MPIs) at midrapidity~\cite{MPI1, MPI2}, and is mostly described by soft quantum chromodynamics (QCD) processes. However, long-range correlations between particle pairs in $\Delta\varphi$, which are usually associated with the development of anisotropic flow, are also observed in the UE region at high multiplicities~\cite{ALICEv2_1}, pointing towards a potential link between the UE and the collective expansion of the medium. 

Using these structural features of the correlation distribution, associated particle production can be partitioned into hard-scattering processes (i.e., near- and away-side regions) and softer processes (i.e., UE region), which can help in exploring the underlying mechanisms responsible for the enhancement of strange hadrons as a function of charged particle multiplicity. Moreover, the correlation distributions can be used to extract the azimuthal widths of the near- and away-side jet peaks~\cite{DeepaPaper, DPaper}, which aid in the study of strange quark production in the context of jet fragmentation.

In this paper, the production of strange quarks in small collision systems is explored by measuring two-particle h--$\Lambda$ azimuthal correlation functions in \pPb collisions at a center-of-mass energy per nucleon pair \snn = 5.02 TeV. The $\Lambda$ baryon was chosen for this analysis as it exhibits a significant enhancement in the $\Lambda$/$\pi$ ratio as a function of multiplicity in \pPb collisions, while still being abundantly produced such that statistically significant differential correlation measurements can be studied. Furthermore, the $\Lambda$ baryon ($m = 1.116$ \GeVmass) has a mass similar to that of the $\phi(1020)$ meson ($m = 1.020$ \GeVmass)\cite{PDG}, making it a good candidate to explore the differences between open (\lmb) and hidden ($\phi$) strangeness production. The choice of collision system was made primarily because the previously measured enhancement is maximal in the multiplicity range that is spanned nearly entirely by \pPb (and \pp) collisions. Additionally, the \pPb system is of interest because it serves as a middle ground between the hard-process-dominated \pp collisions and the softer, collective-dynamics-dominated \PbPb collisions. This allows for the study of the interplay between hard and soft processes in the production of strange hadrons, which is vital to the understanding of the physics behind strangeness enhancement. 

The h--$\Lambda$ correlation functions are used to measure $\Lambda$ production in and out of jets by extracting the per-trigger yields in each kinematic region as a function of $p_{\text{T}}^{\Lambda}$ and event multiplicity. These results are then compared to similar measurements of inclusive charged hadrons (h--h angular correlations), allowing for the study of strangeness enhancement via the per-trigger yield ratio $\Lambda/\text{h}$ ($\approx \Lambda/\pi$) as a function of multiplicity in the context of jet and UE production. To further investigate the fundamental mechanisms responsible for strangeness production, the near- and away-side jet widths are also extracted from the h--\lmb correlation distributions and compared to those obtained from the dihadron sample. Additionally, the differences between open ($|S| \neq 0$, where $|S|$ is the net strangeness) and hidden ($|S| = 0$) strangeness production are explored using previously published h--$\phi(1020)$ angular correlation measurements in \pPb collisions~\cite{ALICE:2024aid}. These measurements are additionally compared with the DPMJET (v3.0-5) Monte Carlo event generator~\cite{DPMJET}, which uses the Dual Parton Model (DPM)~\cite{DualParton} to describe interactions between nuclei, and is capable of simulating pp, \pPb, and \PbPb collisions.

The article is organized in the following way. Section~\ref{sec:experiment} outlines the experimental setup, the data sample, and the data selection criteria. Section ~\ref{sec:analysis} describes the analysis procedure, with Sec.~\ref{sec:systematics} detailing the systematic uncertainties associated with this procedure. Finally, Sec.~\ref{sec:results} presents the results of the analysis, and Sec.~\ref{sec:conclusion} summarizes the conclusions of this study.

\section{Experimental setup}
\label{sec:experiment}

The data used for the measurements presented in this paper were collected with the ALICE detector during the LHC Run 2 data taking campaign (2015--2018). The ALICE detector is composed of a forward muon spectrometer and a central barrel system located within a 0.5 T solenoidal magnetic field oriented along the beam axis. More information about the ALICE detector and its performance can be found in ~\cite{ALICEexperiment1, ALICEexperiment2,ALICEExperiment3}. The measurements presented in this paper mostly rely on data reconstructed with the central barrel detectors, namely the Inner Tracking System (ITS), the Time Projection Chamber (TPC), and the Time of Flight (TOF) detector. The ITS~\cite{ITS} is the innermost detector of ALICE and, during Run 2, was composed of six cylindrical layers of silicon detectors. The ITS was used to reconstruct both primary and secondary vertices, as well as for the tracking of charged particles. The TPC~\cite{TPC} is a gas-filled chamber capable of reconstructing charged particle tracks in three dimensions. Additionally, it allows for charged particle identification via specific energy loss measurements (\dEdx)~\cite{TPCPID}. The TOF~\cite{TOF} detector is a large array of Multi-gap Resistive Plate Chambers (MRPCs) capable of measuring the time of flight of charged particles with a timing resolution of around 60 ps. The TOF detector is used in conjunction with the TPC to improve charged particle identification via time-of-flight measurements. 

Finally, the V0, a detector composed of two arrays of scintillator counters (V0A, V0C) located on either side of the nominal interaction point in the pseudorapidity ranges $2.8 < \eta < 5.1$ (V0A) and $-3.7 < \eta < -1.7$ (V0C)~\cite{V0Detector}, was used for both determining the event activity (multiplicity) of the collisions and providing the trigger for the data acquisition system. 

\subsection{Event selection}
The dataset analyzed in this paper consists of \pPb collisions at \snn = 5.02 TeV, recorded in 2016. The data were collected using a minimum bias trigger, which requires a coincident signal in both the V0A and V0C detectors. Furthermore, each event in this analysis is required to have a primary vertex (PV) within 10 cm of the nominal collision point along the beam axis to ensure uniform central barrel detector performance. Moreover, each event is required to contain at least three reconstructed charged hadrons within midrapidity ($|\eta| < 0.8$) with a $p_{\rm T}$ greater than 0.15 GeV/$c$ to fit the minimum criteria for a correlation measurement. After the event selection, the total number of analyzed events is $\approx 4 \times 10^8$, corresponding to an integrated luminosity of $\approx 2$ nb$^{-1}$~\cite{ALICEpPbCrossSection}.

To measure the correlation distribution dependence on charged particle multiplicity, the selected events are further categorized into three multiplicity classes: 0--20\%, 20--50\%, and 50--80\%. These classes are determined by event activity within the V0A detector (Pb-facing), with 0--20\% corresponding to the 20\% of events with the highest V0A signal. These three classes were chosen to evenly distribute the $\Lambda$ signal in each class to reduce statistical fluctuations in the corresponding correlation distributions. For events generated using DPMJET, the multiplicity classes are determined by the number of charged primary particles produced within the $\eta$ acceptance of the V0A detector.

\subsection{Track selection and $\Lambda$ reconstruction}

To measure the h--$\Lambda$ and h--h correlation functions required for separating jet and UE associated production, three different types of particles are needed: high-momentum trigger particles which serve as a proxy for the jet axis, associated $\Lambda$'s, and associated charged hadrons. All measured pseudorapidities ($\eta$) are referred to the laboratory frame. 

The trigger hadrons are reconstructed in the pseudorapidity range $|\eta| < 0.8$ using information from both the ITS and the TPC. To ensure that the trigger hadron sample contains high-quality tracks, the number of crossed pad rows in the TPC is required to be $\geq 70$ (out of a maximum of 159) and the ratio of the number of crossed rows to the number of findable clusters -- possible assignable clusters to a track -- must be $> 0.8$. In addition, tracks that share more than 40\% of their clusters with other tracks are rejected. In order to select primary particles, the distance of closest approach (DCA) to the PV is required to be $< 2.4$ cm in the transverse plane and $<3.2$ cm along the beam axis. Finally, the goodness-of-fit ($\chi^2$) per cluster must be smaller than 36 for the track fit in the ITS and 4 for the fit in the TPC. All selected trigger hadrons are required to have transverse momentum $4.0 <$ \pt $< 8.0$ GeV/$c$ to ensure that the trigger sample is dominated by jet fragmentation products. The upper bound of 8.0 \GeVc was chosen such that the trigger reconstruction efficiency can be computed using a simulated MC sample (see Section~\ref{sec:analysis}). In this momentum range, the contamination from secondary particles is less than 1\%. The numbers of trigger hadrons in the 0--20\%, 20--50\%, and 50--80\% multiplicity classes are 5.3$\times 10^6$, 4.3$\times 10^6$, and 1.8$\times 10^6$, respectively. Of the total selected events, only around 10\% contain a trigger hadron, with $<$1\% of events containing more than one trigger hadron. All trigger hadrons within an event are used to calculate the correlation functions.

The associated hadrons in the dihadron correlation are also reconstructed in the pseudorapidity range $|\eta| < 0.8$ using selection criteria nearly identical to those of the trigger hadrons with two differences. The first is a much stricter, \pt-dependent DCA selection criterion along the transverse plane of the form DCA$_{xy} < $ $[0.0105 + 0.0350 \times p_{\text{T}}^{-1.1}]$ cm, where \pt is measured in \GeVc. The second is the requirement of a hit in one of the two innermost layers of the ITS. Both of these requirements are imposed to ensure the selection of primary hadrons. All associated hadrons were selected in the momentum ranges $1.5 <$ \pt $< 2.5$ GeV/$c$ (lower \pt) and $2.5 <$ \pt $< 4.0$ \GeVc (higher \pt).

\begin{table}[h]
\centering
\caption{Topological selection criteria applied to $\Lambda$ candidates.}
\begin{tabular}{l c}
\hline
Selection criterion & Value \\
\hline
$|\eta|$ & $< 0.8$ \\
Decay radius (cm) & $> 0.2$ \\
DCA$_{xy}$ of pion track to PV (cm) & $> 0.06$ \\
DCA$_{xy}$ of proton track to PV (cm) & $> 0.06$ \\
DCA$_{xy}$ between daughter tracks (n$\sigma$) & $< 1.5$ \\
cos($\theta_{\rm pointing}$) & $> 0.9$ \\
Invariant mass (\GeV/$c^2$) & $1.102 < M_{\text{p}\pi} < 1.130$ \\
\hline
\end{tabular}
\label{tab:LambdaCuts}
\end{table}

\begin{figure}[h]
\centering
\includegraphics[width=0.49\textwidth]{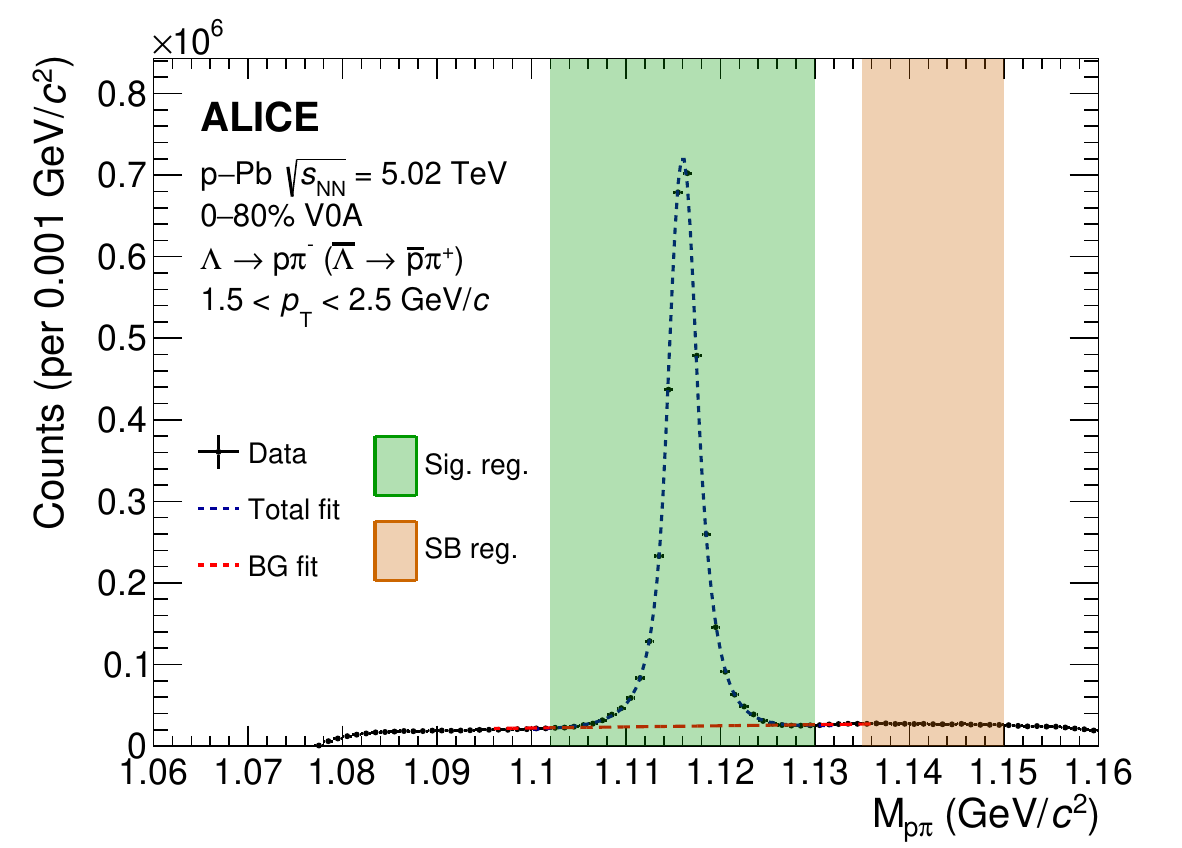}
\includegraphics[width=0.49\textwidth]{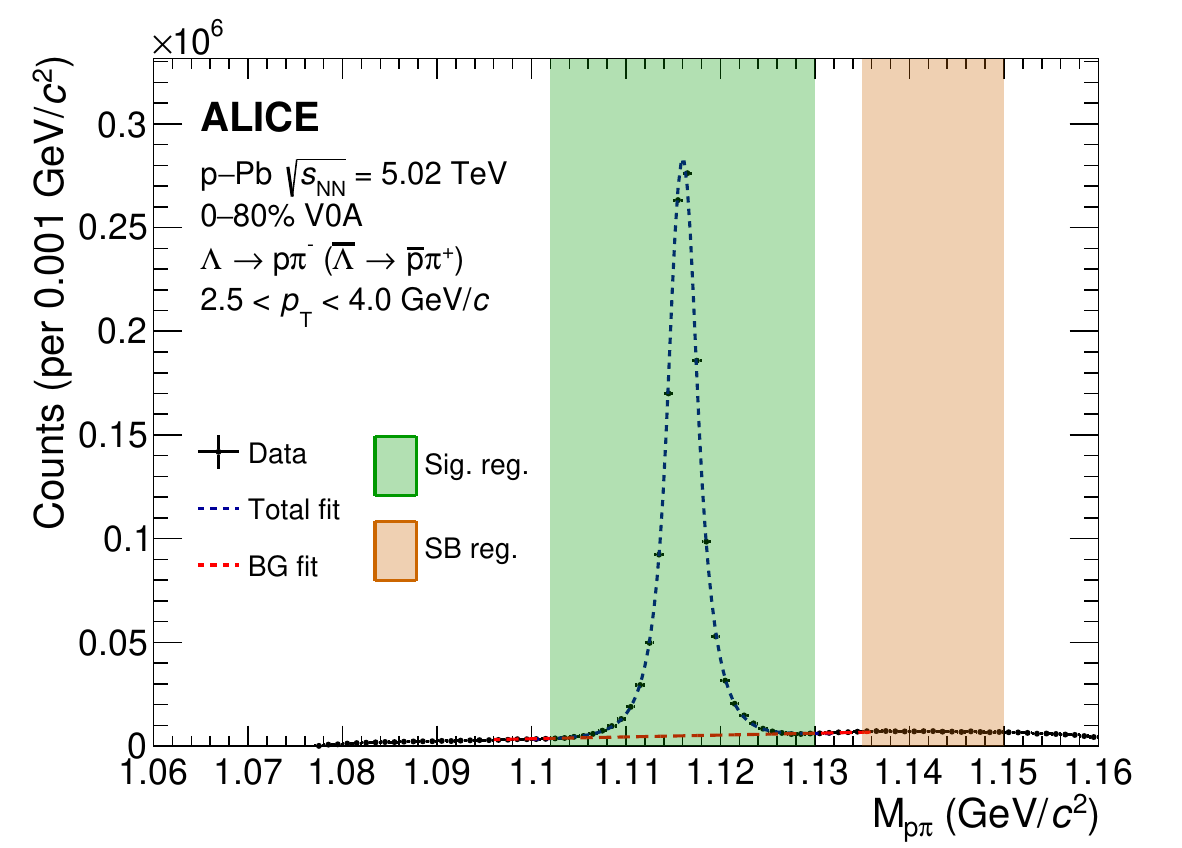}
\caption{Invariant mass distributions of $\Lambda$ candidates in events with a trigger hadron for the lower ($1.5 <$ \pt $<2.5$ \GeVc, left) and higher ($2.5 <$ \pt $< 4.0$ \GeVc, right) associated momentum ranges. A double-sided Crystal Ball + linear background function is used to fit the data. The signal and side-band (SB) regions are highlighted in green and orange, respectively. The SB region is used to subtract the combinatorial background from the correlation distributions.}
\label{fig:LambdaMass}
\end{figure}

The associated $\Lambda$ baryons ($\Lambda + \overline{\Lambda}$) are reconstructed in the pseudorapidity range $|\eta| < 0.8$ via the decay channel $\Lambda \rightarrow \text{p} + \pi^{-}$, which has a branching ratio (BR) of 64.1\%~\cite{PDG}. Following techniques similar to those presented in Refs.~\cite{V0Reco1} and~\cite{V0Reco2}, the $\Lambda$ baryons are reconstructed by exploiting their characteristic V-shaped weak decay topology. The daughter proton and pion tracks are identified via specific energy loss in the TPC and their timing information from the TOF detector. The daughter protons (pions) are required to fall within $\pm2\sigma$ ($\pm3\sigma$) of the expected mean value for the TPC and TOF signal. More details about this procedure can be found in Refs.~\cite{PIDSource,ALICEpPbEnhancement}. If there is no signal in the TOF detector from the daughter track due to the detector's smaller acceptance, only the TPC signal is used for particle identification. The $\Lambda$ daughter track quality requirements are similar to those of the trigger hadron, but the DCA to the PV is required to be $> 0.06$ cm in the transverse plane for both daughter particles. The pairs of identified proton and pion tracks are combined into $\Lambda$ candidates, which are then selected by requiring the invariant mass of the pair to be within 14 MeV/$c^2$ of the known $\Lambda$ mass of 1.116 \GeVmass~\cite{PDG}. To reduce the combinatorial background, further requirements on the topological variables associated with the $\Lambda$ decay are applied. These requirements are summarized in Table ~\ref{tab:LambdaCuts}. The decay radius is the distance between the secondary vertex (the point where the $\Lambda$ decays) and the PV. The DCA$_{xy}$ between the daughter tracks is measured in terms of its resolution $\sigma$. The pointing angle $\theta_{\rm pointing}$ is the angle between the vector defined by the PV and the secondary vertex, and the momentum vector of the $\Lambda$ candidate. The requirements presented in Table~\ref{tab:LambdaCuts} are less strict than in previous analyses~\cite{Lambda1, Lambda2} and are chosen to maximize the $\Lambda$ signal while maintaining a low ($< 10\%$) background level. Similarly to the associated charged hadrons, the $\Lambda$ baryons are selected within the momentum intervals $1.5 <$ \pt $< 2.5$ \GeVc  and $2.5 <$ \pt $< 4.0$ \GeVc.   The multiplicity-integrated invariant mass distributions of $\Lambda$ candidates in events with a trigger hadron for the lower and higher associated momentum ranges are shown in Fig.~\ref{fig:LambdaMass}. To constrain the combinatorial background, the distributions are fitted with the sum of a double-sided Crystal Ball function~\cite{CrystalBall1} and a linear background function. The double-sided Crystal Ball function is chosen to account for the tails in the invariant mass distribution, which exhibit non-Gaussian behavior. The extracted $\Lambda$ signal from these distributions is around $3.2 \times 10^6$ in the lower-\pt range and $1.3 \times 10^6$ in the higher-\pt range. To maximize the available $\Lambda$ signal, the $\Lambda$-baryon sample includes those that have decayed from the $\Xi^-$ and $\Xi^0$ baryons (and their antiparticles), which contributes approximately 15\% to the total $\Lambda$ sample in the given momentum ranges~\cite{Feeddown}. The sample also includes $\Lambda$ baryons which decayed from $\Omega^-$ baryons, which constitute less than 1\% of the total $\Lambda$ sample and are therefore negligible relative to the systematic uncertainties presented in Section~\ref{sec:systematics}. The $\Lambda$-baryon sample from DPMJET also contains both primary and secondary $\Lambda$ baryons, with proportions similar to the data.

\section{Analysis details}
\label{sec:analysis}

\subsection{Angular correlation functions}
The observables used in this analysis are derived from two-particle angular correlation functions, which can be defined in multiple ways as discussed in Ref.~\cite{CorrelationFunction}. The focus of this analysis is the associated particle production in and out of jets, quantified by the per-trigger normalized associated particle yield,
\begin{equation}
    C_{\text{yield}}(\Delta\varphi, \Delta\eta) = \frac{1}{N_{\text{trig}}^{\text{corr}}}\frac{1}{\epsilon_{\text{trig}}\times\epsilon_{\text{assoc}}}B(0,0)\frac{S(\Delta\varphi, \Delta\eta)}{B(\Delta\varphi, \Delta\eta)},
\label{eq:corr_detector}
\end{equation}
where the single-particle efficiency correction factors $1/\epsilon_{\text{trig}}$ and $1/\epsilon_{\text{assoc}}$ are applied for each trigger-associated pair. The angular separations $\Delta\varphi$ and $\Delta\eta$ are measured between a high-momentum charged hadron trigger (h), which serves as a proxy for the jet axis, and a lower momentum associated hadron ($\Lambda$, h). 

\begin{figure}[t]
\makebox[\linewidth][c]{%
\centering
\subfigure{
  \label{fig:a}
  \includegraphics[width=0.35\textwidth]{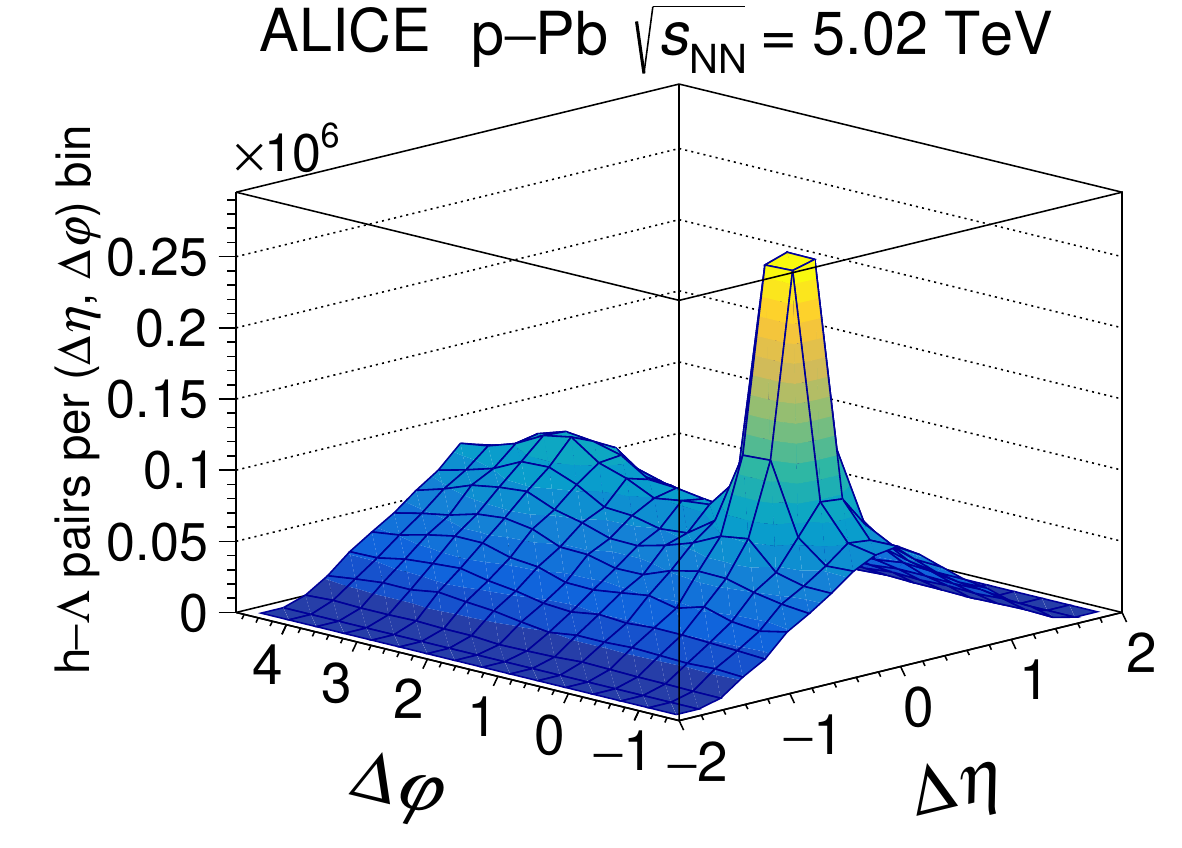}
}
\subfigure{
  \label{fig:b}
  \includegraphics[width=0.35\textwidth]{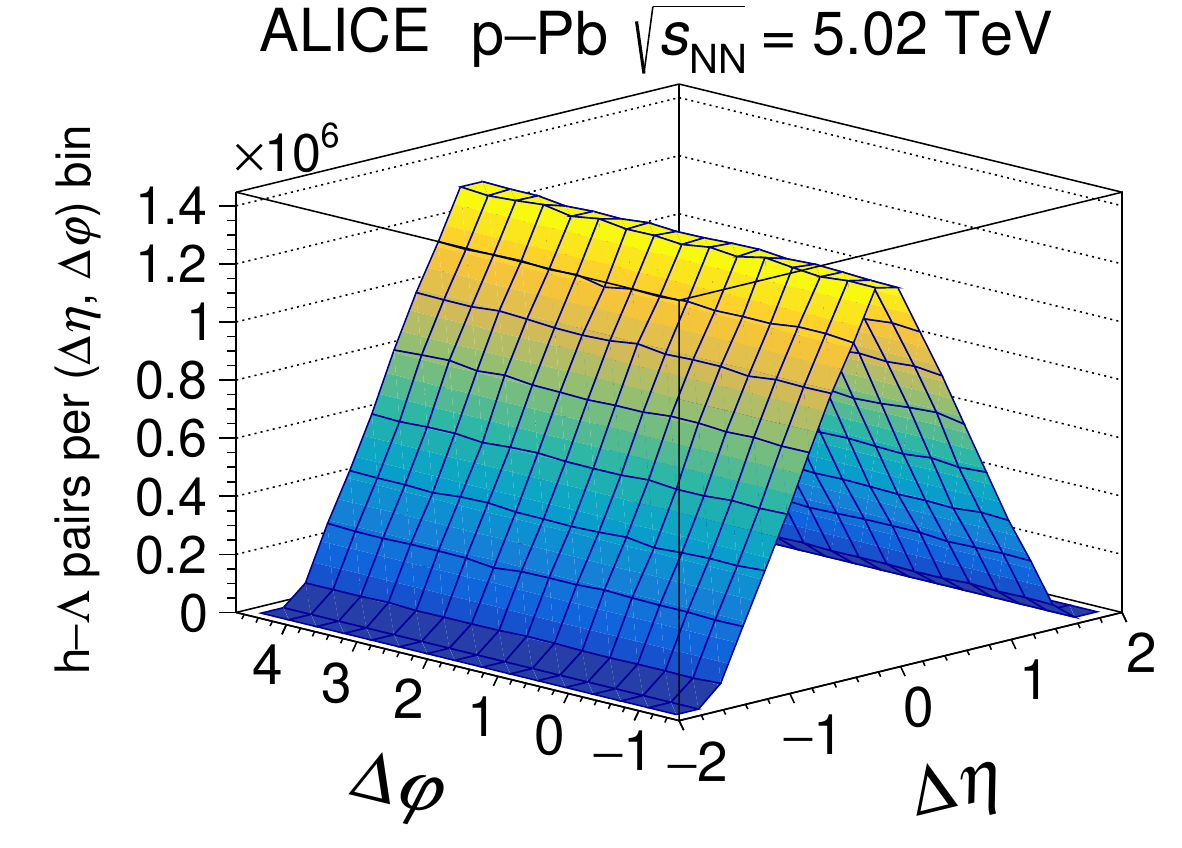}
}
\subfigure{
  \label{fig:c}
  \includegraphics[width=0.35\textwidth]{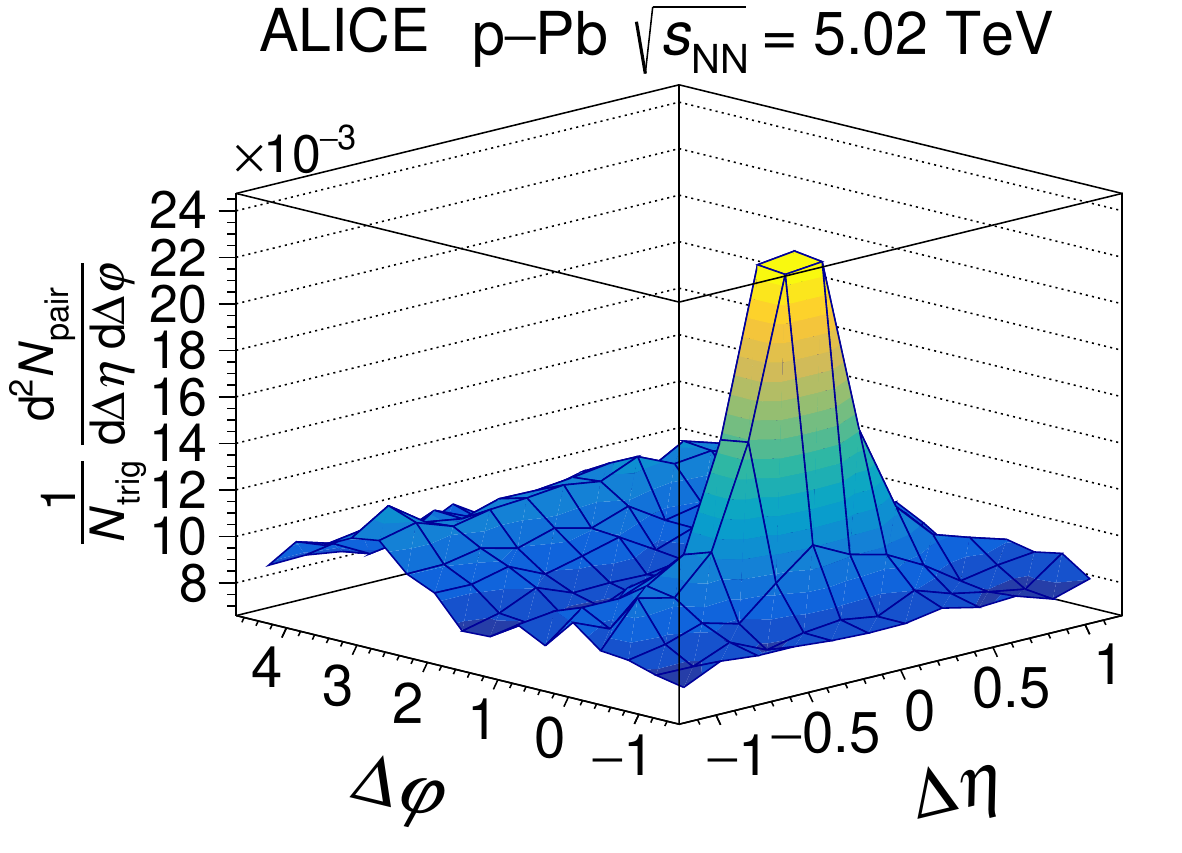}
}
}
\caption{Examples of the h--\lmb single-particle efficiency-corrected same-event distribution $S(\Delta\varphi, \Delta\eta)$ (left), mixed-event distribution $B(\Delta\varphi, \Delta\eta)$ (middle), and fully corrected correlation function $C_{\text{yield}}(\Delta\varphi, \Delta\eta)$ (right), taken in the 20--50\% multiplicity class.}
\label{fig:twod_cor}
\end{figure}

The quantity $S(\Delta\varphi, \Delta\eta)$ is the raw distribution of trigger--associated pairs,
\begin{equation}
S(\Delta\varphi, \Delta\eta)  = \frac{\text{d}^2N_{\text{pair}}}{\text{d}\Delta\varphi\text{d}\Delta\eta}, 
\end{equation}
where $N_\text{pair}$ is the number of trigger--associated pairs produced within the same event with angular separation $(\Delta\varphi, \Delta\eta)$. Both $\epsilon_{\text{trig}}$ and $\epsilon_{\text{assoc}}$ are obtained using a MC sample generated with DPMJET (v3.0.5) ~\cite{DPMJET} and propagated through the ALICE detector using GEANT3 ~\cite{GEANT}. The trigger and associated particle efficiency factors are calculated separately as a function of \pt, $\eta$, and event multiplicity by comparing the reconstructed and generated particle distributions. The weight $1/(\epsilon_{\text{trig}}\times\epsilon_{\text{assoc}})$ is applied for each pair in the raw distribution as it is being filled. The trigger and associated charged hadron efficiencies are around 80\% for all \pt ranges, and the $\Lambda$ efficiency varies from 35\% at the lowest \pt range to 50\% at the highest. A plot of $S(\Delta\varphi, \Delta\eta)$ after the single-particle efficiency corrections for h--\lmb pairs can be seen in the left panel of Fig.~\ref{fig:twod_cor}. The trigger efficiency weight $1/\epsilon_{\text{trig}}$ is also applied to the single-particle trigger hadron distribution in data to obtain $N_{\text{trig}}^{\text{corr}}$, which is the total number of triggers in the event sample.

In Eq.~\ref{eq:corr_detector}, $B(\Delta\varphi, \Delta\eta)$ is the distribution generated by combining trigger and associated particles that are produced in separate events, often called the mixed-event distribution~\cite{MixedEvent1, MixedEvent2, MixedEvent3}. This distribution is also corrected for single-particle efficiency in the same manner as $S(\Delta\varphi, \Delta\eta)$. The factor $B(0, 0)/B(\Delta\varphi, \Delta\eta)$ is used to correct for the finite acceptance along $\eta$ as both the trigger and associated particles are required to be within $|\eta| < 0.8$. As the single particle $\eta$ distributions are uniform in the selected range, the mixed-event distribution has a characteristic triangular shape along $\Delta\eta$, which is purely due to detector geometry as no physical correlations are present. The quantity $B(\Delta\varphi, \Delta\eta)$/$B(0, 0)$ represents the probability that a particle pair with angular separation $(\Delta\varphi, \Delta\eta)$ is found given that the trigger particle is within $|\eta| < 0.8$, and is used to correct for pair acceptance and pair efficiency. A plot of the mixed-event distribution can be seen in the middle panel of Fig.~\ref{fig:twod_cor}.

While Eq.~\ref{eq:corr_detector} is applicable for both the dihadron and h--$\Lambda$ correlation functions, the h--$\Lambda$ case requires additional corrections that are not present in the dihadron case.

First, the topological selection criteria for the $\Lambda$ reconstruction result in a small amount of combinatorial background. To remove it from the h--$\Lambda$ correlation distribution, an additional distribution is made from the same selection criteria, but with the $\Lambda$ mass region shifted to the right such that $1.135 < M_{\text{p}\pi} < 1.150$ GeV/$c^2$ as shown by the orange highlighted ``side-band'' region in Fig.~\ref{fig:LambdaMass}. The correlation distribution in the side-band region is divided by the integral of the invariant mass distribution within the same region. The resulting correlation distribution is then scaled by the integral of the background in the signal region, which is obtained from the linear background fit shown in Fig.~\ref{fig:LambdaMass}. Finally, the side-band correlation distribution is subtracted from the nominal distribution.

After this subtraction, the h--$\Lambda$ correlation distribution is corrected for the branching ratio of the $\Lambda$ decay by scaling the distribution by $1/\text{BR}$, where $\text{BR}$ is the branching ratio. To correct for the $\Lambda$ baryons that are excluded by the nominal mass window, the h--$\Lambda$ distribution is scaled by $1/f_\text{sig}$, where $f_\text{sig}$ is the fraction $\Lambda_{\text{signal}}/\Lambda_{\text{total}}$. In this fraction, $\Lambda_{\text{signal}}$ is the number of $\Lambda$ candidates within the nominal mass window, and $\Lambda_{\text{total}}$ is the total number of $\Lambda$ candidates, both obtained via integrating the invariant mass distribution after removing the combinatorial background. As the invariant mass window for the signal region encompasses nearly the entire $\Lambda$ peak, $f_\text{sig}$ is very close to unity ($\approx 0.98$).

Finally, the correlation distribution is corrected for h--$\Lambda$ pairs that are lost in the Kalman filtering procedure during track reconstruction. Indeed, higher-quality tracks (those reconstructed with more clusters in the TPC and ITS) are favored over lower-quality ones whenever ambiguity arises from clusters being shared between the two tracks. This results in a deficit of reconstructed h--$\Lambda$ pairs at small $\Delta\varphi\Delta\eta$ due to the merging of the tracks of the trigger particle and of the proton produced in the $\Lambda$ decay. This effect, already observed in previous analyses~\cite{TwoTrack1, TwoTrack2}, is more substantial in this case due to: 1) the large decay length of the $\Lambda$, resulting in less detector information to reconstruct the daughter tracks, and 2) the asymmetry in the $\Lambda$ decay products ($\frac{m_{\text{p}}}{m_{\pi}} \approx 7$) causing the daughter proton to occupy a similar phase space as the $\Lambda$ (i.e., they share similar $\varphi$ and $\eta$). Thus, if a trigger particle and a $\Lambda$ decay proton are reconstructed as a single track, a h--$\Lambda$ pair will be lost at small $\Delta\varphi\Delta\eta$. The magnitude of this effect is \pt dependent and varies from up to 10\% in the $1.5 < p_{\text{T}} < 2.5$ GeV/$c$ range and up to 20\% in the $2.5 < p_{\text{T}} < 4.0$ GeV/$c$ range. Following a procedure similar to that in Ref.~\cite{TwoTrack2}, this effect is corrected for via scaling the correlation distribution by
\begin{equation}
    \upsilon_{\text{pair}}(\Delta\varphi, \Delta\eta) = \frac{C_{\text{reco}}^{\text{tag}}(\Delta\varphi, \Delta\eta)}{C_{\text{gen}}(\Delta\varphi, \Delta\eta)},
\label{eq:paircorr}
\end{equation}
where $C_{\text{reco}}^{\text{tag}}$ is the efficiency-corrected correlation distribution calculated in MC using reconstructed trigger hadrons and $\Lambda$ candidates with the same selection criteria as in data. Additionally, the reconstructed $\Lambda$ candidate is required to have a corresponding generated $\Lambda$, which is used for all calculations involving kinematic quantities. This removes any extraneous effects due to combinatorial background or detector resolution. In Eq.~\ref{eq:paircorr}, $C_{\text{gen}}$ is the correlation distribution calculated in MC using generated trigger hadrons and $\Lambda$ candidates. The template $\upsilon_{\text{pair}}(\Delta\varphi, \Delta\eta)$ is applied for each associated \pt interval in this analysis. This template has been verified to be independent of event multiplicity and choice of MC generator for this analysis. This effect has also been observed in the dihadron case; however, the effect is found to have a much smaller influence on the results ($<$ 1\%) and is thus neglected in this analysis.

A plot of the fully corrected distribution $C_{\text{yield}}(\Delta\varphi, \Delta\eta)$ for h--\lmb pairs can be seen in the right panel of Fig.~\ref{fig:twod_cor}. The peaks observed around $\Delta\varphi = 0$ and $\Delta\varphi = \pi$ in the fully corrected correlation function define the aforementioned near- and away-side regions, respectively, which lie on top of the jet-independent UE. To minimize the statistical fluctuations in the h--$\Lambda$ correlation functions, all two-dimensional correlation functions in this analysis are projected onto $\Delta\varphi$ within the range $-1.2 < \Delta\eta < 1.2$ to obtain the one-dimensional azimuthal distribution $\frac{\text{d}N_{\text{pair}}}{\text{d}\Delta\varphi}$.

\subsection{Underlying event fit}
\label{sec:uefit}

To extract the near- and away-side widths and the per-trigger yields in each kinematic region from the azimuthal correlation distributions, the underlying event contribution must be quantified. Previous measurements from ALICE indicate the presence of a non-zero elliptic flow ($v_{2}$) contribution from both the trigger and associated $\Lambda$ (h) particles ~\cite{ALICEv2_1}, resulting in an underlying event contribution that is not flat along $\Delta\varphi$. To account for this, the underlying event fitting function
\begin{equation}
U(\Delta\varphi) = A(1 + 2v_{2}^{\text{trig}}v_{2}^{\text{assoc}}\cos(2\Delta\varphi))
\label{eq:ue_v2}
\end{equation}
is used. This function is fit to the $\Delta\varphi$ distribution in the regions $(-\frac{\pi}{2}, -\frac{\pi}{4}) \cup (\frac{\pi}{4}, \frac{5\pi}{8}) \cup (\frac{11\pi}{8}, \frac{3\pi}{2})$, which are assumed to have little-to-no jet contribution. The values of $v_{2}^{\text{trig}}$ and $v_{2}^{\text{assoc}}$ are taken for each associated \pt range as a \pt-weighted average of the ALICE measurements (either charged hadron or $\Lambda$ from ~\cite{ALICEv2_2}) and are fixed during the fitting procedure, while the pedestal $A$ is allowed to vary. Cross-checks were performed at  $|\Delta\eta| > 1.2$ to verify that the weighted $v_{2}$ values are consistent with the data from this analysis. For these cross-checks, the resulting $\Delta\varphi$ distributions in this $\Delta\eta$ range were fit using Eq.~\ref{eq:ue_v2} with the $v_{2}$ values fixed to the weighted ones, and the fits were found to be consistent with the data.

\subsection{Yield and width extraction}

 After obtaining the underlying event fit $U(\Delta\varphi)$, the associated particle yields in the jet-like and UE regions are extracted using
\begin{equation}
    Y_{\text{near}} = \int_{-\pi/2}^{\pi/2} (\frac{\text{d}N_{\text{pair}}}{\text{d}\Delta\varphi}- U(\Delta\varphi))\text{d}\Delta\varphi,  \  \ Y_{\text{away}}  = \int_{\pi/2}^{3\pi/2} (\frac{\text{d}N_{\text{pair}}}{\text{d}\Delta\varphi}- U(\Delta\varphi))\text{d}\Delta\varphi
    \label{eq:jet_yields}
\end{equation}
and
\begin{equation}
    Y_{\text{UE}} = \int_{-\pi/2}^{3\pi/2} U(\Delta\varphi)\text{d}\Delta\varphi,
\label{eq:ue_yield}
\end{equation}
where the subscripts near, away and UE refer to the near-side, away-side, and underlying event regions, respectively.

In order to quantify the widths of the near- and away-side peak regions, the $\Delta\varphi$ distributions are fit using the function
\begin{equation}
    F(\Delta\varphi) = U(\Delta\varphi) + \frac{e^{\kappa_{\text{near}}\text{cos}(\Delta\varphi - \mu_{\text{near}})}}{2\pi I_0(\kappa_{\text{near}})} + \frac{e^{\kappa_{\text{away}}\text{cos}(\Delta\varphi - \mu_{\text{away}})}}{2\pi I_0(\kappa_{\text{away}})},
\label{eq:fullfit}
\end{equation}
which is composed of two von Mises functions describing the near- and away-side peaks. Von Mises functions~\cite{VonMises} are the circular analogs of Gaussian distributions and provide the best fit to the $2\pi$-periodic $\Delta\varphi$ distributions for this analysis. The quantities $\kappa_{\text{near}}$ and $\kappa_{\text{away}}$ are a measure of the collimation of the near- and away-side peaks, respectively, and $I_{0}$ is the zeroth-order modified Bessel function. The underlying event fit $U(\Delta\varphi)$ is fixed to the function obtained from the fits described in the previous section. Due to symmetry considerations, the means $\mu_{\text{near}}$ and  $\mu_{\text{away}}$ are also fixed to $0$ and  $\pi$, respectively. The widths of the peaks are then quantified via
\begin{equation}
    \sigma_{\text{near,away}} = \sqrt{-2\ln\frac{I_1(\kappa_{\text{near,away}})}{I_0(\kappa_{\text{near,away}})}},
\label{eq:width}
\end{equation}
where $I_0$ and $I_1$ are the zeroth- and first-order modified Bessel functions, respectively.

\section{Systematic uncertainties}
\label{sec:systematics}

The primary observables in this analysis ($\Delta\varphi$ distributions, yields in each region, near- and away-side widths) could all be affected by systematic uncertainties that arise from the various steps in the analysis procedure. These include the trigger and associated hadron track selection, particle identification for the $\Lambda$ daughter tracks, the topological selections applied to the $\Lambda$ candidates, the choice of the signal and side-band invariant mass windows for the h--$\Lambda$ distributions, the estimation of the UE, and the fitting procedure used to extract the widths. The uncertainties corresponding to each of these sources are estimated separately by varying the corresponding selection criteria, or by using an alternative to the nominal approach in the case of the UE estimation and fitting procedure. For each variation of a selection criterion, the effect on the extracted yields ($Y_{\text{near}}, Y_{\text{away}}, Y_{\text{UE}}$) and near- and away-side widths ($\sigma_{\text{near}}$, $\sigma_{\text{away}}$) is determined by repeating the analysis procedure with the modified analysis configuration. To determine whether a given variation results in a statistically significant deviation from the nominal value, a check is performed following the procedure discussed in Ref.~\cite{BarlowCheck}. If a variation results in a deviation from the nominal value that falls below the threshold set in this analysis at $1\sigma$ (where $\sigma$ is computed as specified in Ref.~\cite{BarlowCheck}), the variation is considered to be statistically consistent with the nominal value and is thus excluded from the systematic uncertainty calculation. Summaries of the primary sources of systematic uncertainty for each of the observables obtained from the h--$\Lambda$ correlation distributions in the most central (0--20\%) and least central (50--80\%) multiplicity classes are given in Tables~\ref{tab:systematics_lambda_cent} and~\ref{tab:systematics_lambda_periph}, respectively. The total systematic uncertainty is obtained by summing the individual contributions in quadrature. For the dihadron distributions, only the systematic uncertainties associated with tracking, UE estimation, and the fitting procedure are considered.

\subsection{Selection criteria}

The systematic uncertainty associated with the topological selection criteria for $\Lambda$ reconstruction has been studied in great detail in previous analyses using the same dataset and similar selection criteria~\cite{ALICEpPbEnhancement, Lambda2}. It is found to be multiplicity independent and varies from 3.2\% in the lowest associated \pt interval to 3.0\% in the highest interval. The biases in the near- and away-side widths that may be introduced by varying the topological selection criteria are considered by randomly and independently varying the value of the correlation function in each $\Delta\varphi$ bin by $\pm 1\sigma$, where $\sigma$ is the systematic error associated with the topological selection criteria from Refs.~\cite{ALICEpPbEnhancement, Lambda2}. The widths are then extracted using the nominal procedure. The resulting variation in the widths is found to be 3.1\% in the near-side and 6.1\% in the away-side, and these values are taken as the systematic uncertainty associated with the topological selection criteria for both the near- and away-side widths. 

The uncertainty due to the $\Lambda$ daughter proton and pion identification using the TPC and TOF signals is estimated by varying the allowed deviation from the expected signal in the TPC and TOF (see Section~\ref{sec:experiment} for more details). These variations include both a more strict requirement ($1.3\sigma$ for protons, $1.8\sigma$ for pions) and a looser requirement ($2.8\sigma$ for protons, $4.2\sigma$ for pions). The variations are found to have a small effect on the $\Delta\varphi$ distributions and per-trigger yields (0.5--2\% across all \pt and multiplicity class); however, the effect on the near- and away-side widths is found to be much larger (up to 4\% on the near-side and 10.8\% on the away-side). Requiring a signal for both the proton and pion tracks in the TOF detector was also considered as a possible variation, but the resulting effect on all observables was found to be within the Barlow threshold ($< 1\sigma$) and thus is not included in the systematic uncertainty.

Possible biases arising from the choice of invariant mass regions (signal, side-band) for the $\Lambda$ candidates are considered by varying these regions from their nominal values. The signal region (green highlighted area in Fig.~\ref{fig:LambdaMass}) is varied from its default value to the more narrow regions $1.108 < M_{\text{p}\pi} < 1.124$ \GeVmass and $1.112 < M_{\text{p}\pi} < 1.120$ \GeVmass. These variations result in deviations of 0.4--1.1\% from the nominal $\Delta\varphi$ distributions and per-trigger yields. The resulting deviations from the default values of the extracted near- and away-side widths are larger, ranging from 1\% to 3.9\%. Wider invariant mass signal regions were also considered, but as the original region captures nearly all of the \lmb signal these variations were found to have a negligible effect on all observables. The side-band region (orange highlighted area in Fig.~\ref{fig:LambdaMass}) is varied in three ways. The first is by narrowing the region to $1.135 < M_{\text{p}\pi} < 1.145$ \GeVmass, the second by shifting the region to $1.140 < M_{\text{p}\pi} < 1.145$ \GeVmass while maintaining its nominal width, and the third is by shifting the region to $1.086 < M_{\text{p}\pi} < 1.098$ \GeVmass, which lies on the left side of the $\Lambda$ peak in the invariant mass distribution. The resulting deviations from the nominal $\Delta\varphi$ distributions and per-trigger yields are found to be 0.3--1.7\%, and the deviations from the original extracted widths are found to be 1.0--5.4\% for the near-side and up to 10\% for the away-side. 

The systematic uncertainty for the material budget has also been studied in detail in previous ALICE analyses of the same dataset and nominal requirements~\cite{ALICEpPbEnhancement}. In the case of the $\Lambda$, this value is found to be multiplicity independent and amounts to 0.6\% in the lowest \pt interval and to 1.1\% in the highest interval. These values are taken as the systematic uncertainty for both the h--$\Lambda$ $\Delta\varphi$ distributions and the per-trigger yields. Furthermore, the systematic uncertainty for the material budget on all observables obtained from the dihadron correlations is taken to be zero as the material budget uncertainties associated with charged hadrons across the entire \pt range reported in~\cite{ALICEpPbEnhancement} are found to be negligible.

\subsection{Underlying event estimation}
To estimate the systematic uncertainty on the yields associated with the determination of the underlying event contribution, the following UE variations are considered. First, the nominal UE fit function $U(\Delta\varphi)$ is fit in the more restricted region $(-\frac{\pi}{2}, -\frac{3\pi}{8}) \cup (\frac{3\pi}{8}, \frac{5\pi}{8}) \cup (\frac{11\pi}{8}, \frac{3\pi}{2})$ to further minimize the contribution from the jet components. Second, to test the effects of a flat UE assumption, constant average functions $U_{1}$ and $U_{2}$ are used instead of $U(\Delta\varphi)$, with the $U_{1}$ average taken in the nominal region $(-\frac{\pi}{2}, -\frac{\pi}{4}) \cup (\frac{\pi}{4}, \frac{5\pi}{8}) \cup (\frac{11\pi}{8}, \frac{3\pi}{2})$ and the $U_{2}$ average taken in the more restricted region$ (-\frac{\pi}{2}, -\frac{3\pi}{8}) \cup (\frac{3\pi}{8}, \frac{5\pi}{8}) \cup (\frac{11\pi}{8}, \frac{3\pi}{2})$. Finally, as negative contributions to the near- and away-side yields are unphysical, the function
\begin{equation}
    U_{\text{pos}}(\Delta\varphi) = \begin{cases}
    U(\Delta\varphi) & \text{if } \frac{\text{d}N_{\text{pair}}}{\text{d}\Delta\varphi} - U(\Delta\varphi) > 0 \\
    \frac{\text{d}N_{\text{pair}}}{\text{d}\Delta\varphi} & \text{otherwise} 
    \end{cases}
\end{equation}
is used, where $U(\Delta\varphi)$ is the UE function obtained by the nominal fitting procedure. The resulting uncertainties in the extracted yields due to these variations are found to vary between 4\% and 10\% for the near-side, 3--10\% for the away-side, and 1--2\% for the UE. Extracting the yields by fitting the entire $\Delta\varphi$ distribution with both a von Mises-based and a Gaussian-based function was also considered, but the results were found to be statistically consistent with the nominal procedure and are thus excluded from the final systematic uncertainty.

\subsection{Full fitting function}
Obtaining the widths of the near- and away-side peaks from the $\Delta\varphi$ distributions requires a full fit of the distribution with a fit function $F(\Delta\varphi)$ that well describes both the near- and away-side peak regions and the UE. The nominal choice of the components of $F(\Delta\varphi)$ (Eq.~\ref{eq:fullfit}) was made to maximize the fit stability across all pair combinations, \pt ranges, and multiplicity classes. However, several other fit functions for extracting the widths are considered in this analysis. The first alternative parameterization replaces the von Mises components of $F(\Delta\varphi)$  with Gaussian functions, which is the more standard procedure for extracting the widths of the jet-like regions. Two additional Gaussian functions are also added (with means fixed at $\pi-2\pi$ and $0 + 2\pi$) to better reflect the periodic nature of the data. The extracted widths are then obtained directly from the fitted Gaussian functions via their $\sigma$ parameters. The widths obtained from this variation are found to be systematically lower than those obtained from the nominal procedure by 1--3\% (near-side) and 4--7\% (away-side) across all pair combinations, \pt ranges, and multiplicity classes. The other two variations of $F(\Delta\varphi)$ use the constant average functions $U_{1}$ and $U_{2}$ from above in lieu of the nominal $U(\Delta\varphi)$ term (while maintaining the von Mises components). The widths extracted using $U_1$ are again found to be systematically lower than from the nominal technique, with deviations of 1--2\% (near-side) and 4--10\% (away-side) across all pair combinations, \pt ranges, and multiplicity classes. Using $U_{2}$ results in less deviation from the nominal technique (1--2\% for the near-side, 3--6\% for the away-side).

\begin{table}[h]
\centering
\caption{Summary of the systematic uncertainties for the h--$\Lambda$ $\Delta\varphi$ distribution, the extracted yields in each region, and the near- and away-side peak widths in the 0--20\% multiplicity class with $2.5 < p_{\text{T, assoc}}^{\Lambda} < 4.0$ \GeVc.}
\label{tab:systematics_lambda_cent}
\begin{tabular}{l c c c c c c}
\hline
Source & $\Delta\varphi$ dist. & $Y_{\text{near}}$ & $Y_{\text{away}}$ & $Y_{\text{UE}}$ & $\sigma_{\text{near}}$ & $\sigma_{\text{away}}$ \\
\hline
$\Lambda$ topological selection & 3.0\% & 3.0\% & 3.0\% & 3.0\% & 3.1\% & 6.1\% \\ 
$\Lambda$ daughter PID & 0.8\% & 0.9\% & 0.9\% & 0.8\% & 2.2\% & 2.0\% \\
$\Lambda$ signal mass window & 0.4\% & 0.7\% & 0.5\% & 0.4\% & 3.4\% & 0.5\%\\
$\Lambda$ side-band mass window & 0.4\% & 0.4\% & 0.3\% & 0.4\% & 1.7\% & 1.0\%\\
Material budget & 0.6\% & 0.6\% & 0.6\% & 0.6\% & -- & -- \\
UE estimation & -- & 9.1\% & 10.4\% & 1.4\% & -- & -- \\
Full fit routine & -- & -- & -- & -- & 3.5\% & 9.0\% \\
\hline
Total& 3.2\% & 9.7\% & 10.9\% & 3.3\% & 6.4\% & 11.1\% \\
\hline
\end{tabular}
\end{table}

\begin{table}[h]
\centering
\caption{Summary of the systematic uncertainties for the h--$\Lambda$ $\Delta\varphi$ distribution, the extracted yields in each region, and the near- and away-side peak widths in the 50--80\% multiplicity class with $2.5 < p_{\text{T, assoc}}^{\Lambda} < 4.0$ \GeVc.}
\label{tab:systematics_lambda_periph}
\begin{tabular}{l c c c c c c}
\hline
Source & $\Delta\varphi$ dist. & $Y_{\text{near}}$ & $Y_{\text{away}}$ & $Y_{\text{UE}}$ & $\sigma_{\text{near}}$ & $\sigma_{\text{away}}$ \\
\hline
$\Lambda$ topological selection & 3.0\% & 3.0\% & 3.0\% & 3.0\% & 3.1\% & 6.1\% \\ 
$\Lambda$ daughter PID & 2.0\% & 2.1\% & 2.1\% & 1.7\% & 1.4\% & 10.8\% \\
$\Lambda$ signal mass window & 1.1\% & 1.3\% & 1.0\% & 0.7\% & 2.5\% & 3.9\%\\
$\Lambda$ side-band mass window & 1.6\% & 1.7\% & 1.7\% & 1.4\% & 0.6\% & 5.7\%\\
Material budget & 1.1\% & 1.1\% & 1.1\% & 1.1\% & -- & -- \\
UE estimation & -- & 2.9\% & 4.9\% & 0.7\% & -- & -- \\
Full fit routine & -- & -- & -- & -- & 1.5\% & 5.4\% \\
\hline
Total& 4.3\% & 5.2\% & 6.5\% & 3.9\% & 4.6\% & 15.2\% \\
\hline
\end{tabular}
\end{table}

\section{Results}
\label{sec:results}

\subsection{Per-trigger $\Delta\varphi$ distributions}

\begin{figure}[h!]
\centering
\includegraphics[width=0.83\textwidth]{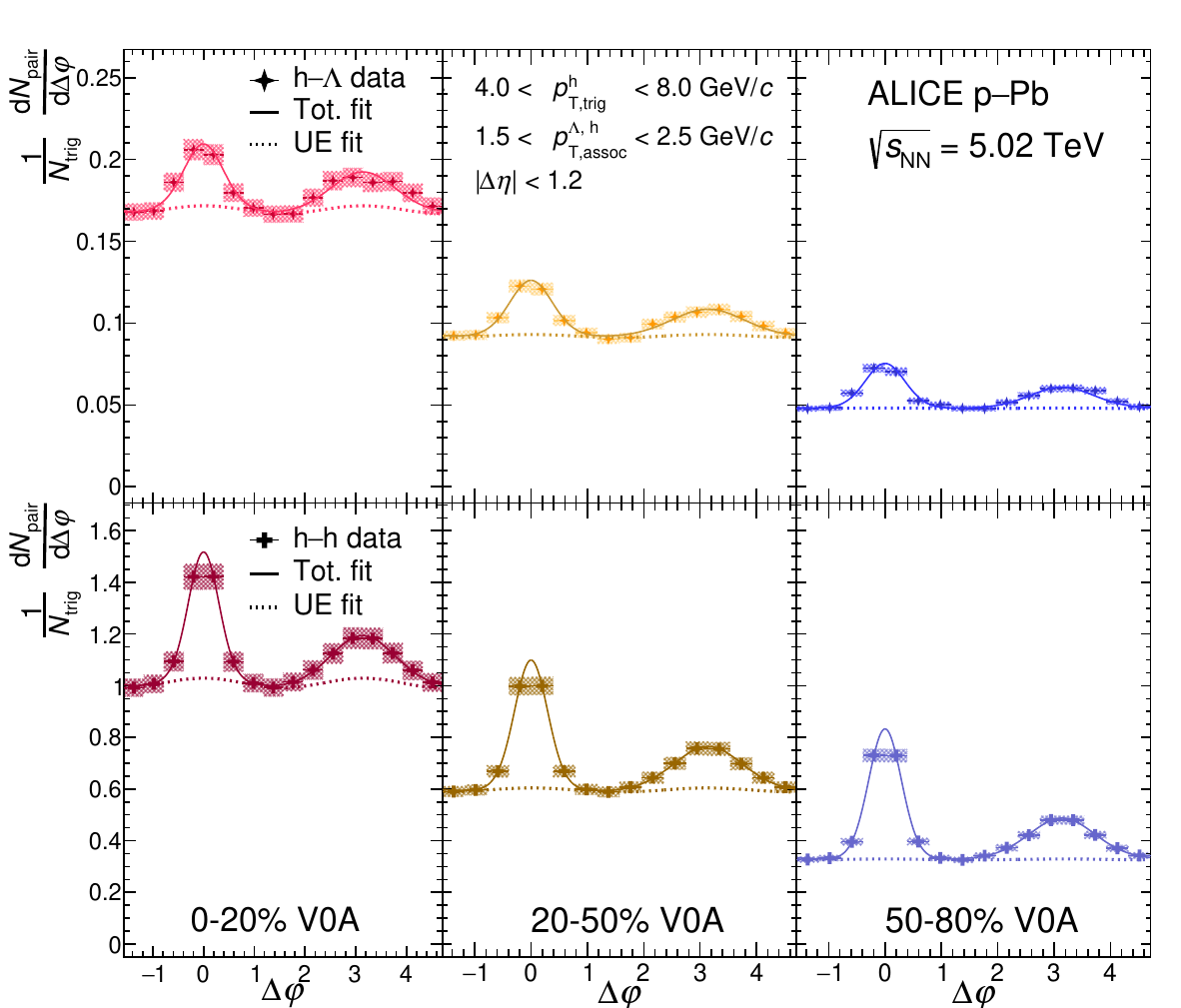}
\caption{The h--$\Lambda$ (top) and h--h (bottom) $\Delta\varphi$ distributions for each multiplicity class with $4.0 < p_{\text{T,trig}} < 8.0$ \GeVc and $1.5 < p_{\text{T,assoc}} < 2.5$ \GeVc, with statistical (systematic) uncertainties shown as vertical lines (shaded boxes). The multiplicity classes are plotted from the most central (left) to the least central (right). The total fit to the data (Eq.~\ref{eq:fullfit}) is shown as a solid line, and the UE estimate with the $v_{2}$ assumption is shown as a dashed line.}
\label{fig:dphi_final_lowpt}
\end{figure}

\begin{figure}[h!]
\centering
\includegraphics[width=0.83\textwidth]{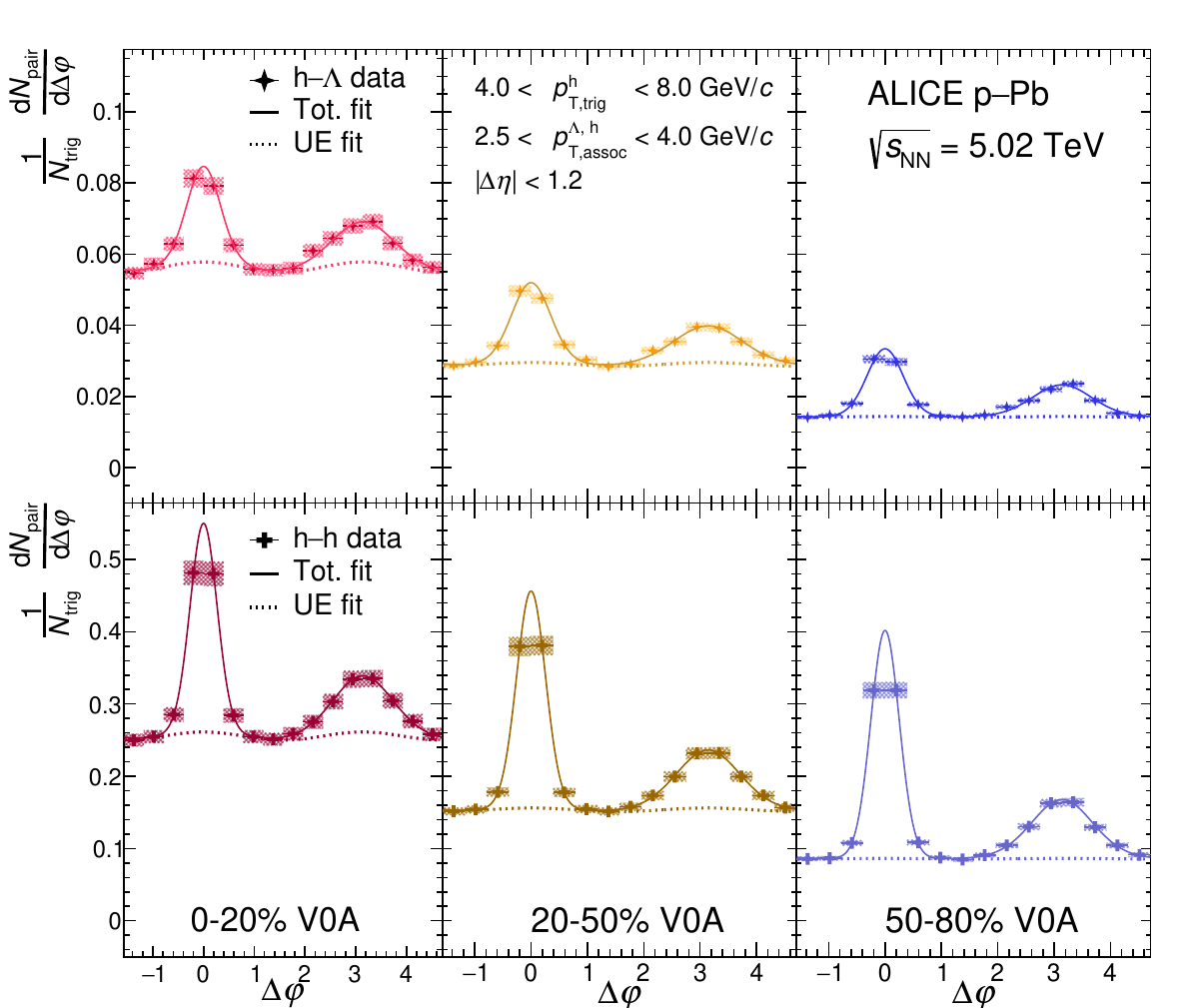}
\caption{The h--$\Lambda$ (top) and h--h (bottom) $\Delta\varphi$ distributions for each multiplicity class with $4.0 < p_{\text{T,trig}} < 8.0$ \GeVc and $2.5 < p_{\text{T,assoc}} < 4.0$ \GeVc, with statistical (systematic) uncertainties shown as vertical lines (shaded boxes). The multiplicity classes are plotted from the most central (left) to the least central (right). The total fit to the data (Eq.~\ref{eq:fullfit}) is shown as a solid line, and the UE estimate with the $v_{2}$ assumption is shown as a dashed line.}
\label{fig:dphi_final_highpt}
\end{figure}

The final per-trigger h--$\Lambda$ and h--h $\Delta\varphi$ distributions are obtained from the 0--20\%, 20--50\%, and 50--80\% V0A multiplicity classes in the trigger momentum range $4.0 < p_{\text{T,trig}}^{\text{h}} < 8.0$ \GeVc and associated momentum ranges $1.5 < p_{\text{T, assoc}}^{\text{h,}\Lambda} < 2.5$ and $2.5 < p_{\text{T, assoc}}^{\text{h,}\Lambda} < 4.0$ \GeVc. These associated \pt ranges are referred to as the ``lower'' and ``higher'' range, respectively. The h--\lmb and h--h correlations and their corresponding fits in the lower and higher associated \pt ranges can be seen in Figs.~\ref{fig:dphi_final_lowpt} and~\ref{fig:dphi_final_highpt}, respectively. The y-axis scale of the plots is tuned to allow a direct comparison of the correlations in the different centrality classes and to emphasize the relative contribution to each distribution from the UE. The UE fit function is calculated using the procedure described in Section~\ref{sec:uefit}. 

In the lower associated momentum range, the UE pedestals for both the h--\lmb and h--h distributions are found to increase by around a factor of 3 from the lowest to the highest multiplicity class (0.05 to 0.17 in the h--\lmb case, 0.35 to 1 in the dihadron case). The higher associated momentum range exhibits a similar increase in the UE pedestal with increasing multiplicity, but the h--\lmb pedestal increases by a factor of 4 instead of 3. The UE pedestal is also found to be higher in the lower associated \pt range than in the higher range by around a factor of 3 in the h--\lmb case and 4 in the h--h case for each multiplicity class. Furthermore, the relative contribution to the total distribution from the UE with respect to jet production increases with increasing multiplicity in both \pt intervals, with the lower-\pt interval exhibiting a larger relative contribution from the UE than the higher-\pt interval across all multiplicity classes. The latter observation suggests that associated production in the UE region is truly ``softer'' than production in the near- and away-side regions, as expected.

\subsection{Per-trigger yields and yield ratios}

To compare with previous results~\cite{ALICEpPbEnhancement,ALICEppEnhancement}, the V0A multiplicity classes have been converted to charged particle multiplicity by computing the charged particle pseudorapidity density $\langle\text{\dndeta}\rangle$ for each multiplicity class in events with a trigger hadron, considering only charged hadrons with $|\eta| < 0.5$ and \pt $> 0.15$ \GeVc. The computation was performed using techniques similar to those reported in Ref.~\cite{ALICEPseudo}. The values of $\langle\text{\dndeta}\rangle$ for each multiplicity class in minimum bias (MB) and triggered events can be seen in Table~\ref{tab:dndeta}. For the MB events, the values are also taken from Ref.~\cite{ALICEpPbEnhancement}. As the high-\pt trigger hadron requirement biases the event sample towards higher multiplicities, the values of $\langle\text{\dndeta}\rangle$ are higher in the triggered events than in the MB events by around $+6$ for each multiplicity class.

\begin{table}
\centering
\caption{The values of $\langle\text{\dndeta}\rangle_{|\eta| < 0.5}$ and the corresponding total uncertainties for each multiplicity class in minimum bias (MB) events and events with a trigger hadron with \pt $> 4$ \GeVc.}
\begin{tabular}{l c c }
\hline \\ [-2.5ex]
Mult. class & $\langle\text{\dndeta}\rangle_{|\eta < 0.5}^{\text{MB}}$  & $\langle\text{\dndeta}\rangle_{|\eta| < 0.5}^{p_{\text{T, trig}} > 4 \text{\GeVc}}$ \\
\hline
0--20\% & $35.6 \pm 0.9$ & $42.4 \pm 0.9$ \\
20--50\% & $21.5 \pm 0.5$ & $27.6 \pm 0.5$ \\
50--80\% & $12.0 \pm 0.3$ & $17.7 \pm 0.4$ \\
\hline
\end{tabular}
\label{tab:dndeta}
\end{table}

\begin{figure}[h!]
\centering
\includegraphics[width=\textwidth]{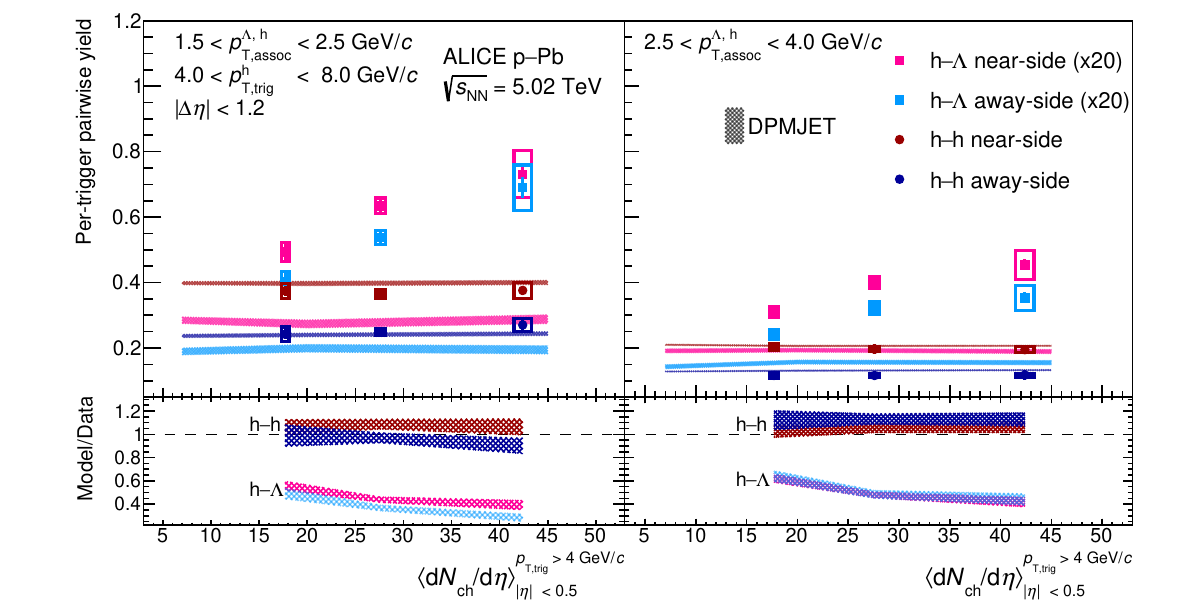}
\caption{The per-trigger pair-wise yields $Y_{\text{near}}, Y_{\text{away}}$ as a function of charged particle multiplicity for the h--$\Lambda$ (square markers) and h--h (circle markers) correlations in the lower (left) and higher (right) associated \pt ranges. The statistical (systematic) uncertainties are shown as vertical lines (boxes). The same yields predicted by DPMJET are shown as shaded bands, with the width of the band representing the uncertainty of the model. The h--\lmb yields (data and DPMJET) have been scaled by a factor of 20 to increase visibility. The ratios of the model to the data are also shown as shaded bands in the bottom panels, where the width of the band represents the uncertainty of the ratio. A dashed line is drawn at unity for reference.}
\label{fig:pairwise_yield}
\end{figure}

The per-trigger yields in the near- and away-side regions of the $\Delta\varphi$ distributions ($Y_{\text{near}}$, $Y_{\text{away}}$) are shown in each associated \pt range as a function of multiplicity for both the h--$\Lambda$ and dihadron correlations in Fig.~\ref{fig:pairwise_yield}. The same yields as predicted by DPMJET are also shown, with their ratios to the data presented in the bottom panels. A dashed line at unity is drawn to help better visualize the deviations between data and the model. The h--\lmb yields measured in data and predicted by DPMJET are scaled by a factor of 20 to increase visibility.

Across both associated \pt ranges, the h--$\Lambda$ yields show a substantial increase with respect to multiplicity for both the near- and away-side regions. This is in stark contrast to the dihadron yields, which exhibit no significant increase as a function of multiplicity in both associated \pt ranges. The increase can be quantified by calculating the percent change in the per-trigger yields from the lowest to the highest multiplicity class, which is shown for each momentum range in Table~\ref{tab:percent_increase}. The uncertainties reported are calculated using both the systematic and statistical uncertainties summed in quadrature, and the significance is obtained by calculating the deviation in the percent change from zero in terms of the total uncertainty. The significances obtained from the h--$\Lambda$ yields across both regions and \pt ranges are all close to $3\sigma$. However, the dihadron yields see no statistically significant increase in both regions across both momentum ranges. In the lower-\pt range, for both the h--\lmb and h--h cases, the percent changes in the away-side yields are systematically higher than the changes in the near-side yields. However, the significance of the difference in the percent change is only around $0.8\sigma$ for both cases, indicating that the observed differences between the near- and away-side increases in this \pt interval are not statistically significant. 

\begin{table}
\centering
\caption{The percent change in the per-trigger yields from the lowest to the highest multiplicity class in the lower and higher associated momentum ranges. The uncertainties reported are obtained using the systematic and statistical uncertainties summed in quadrature. The reported significance is the number of standard deviations away from zero percent change.}
\begin{tabular}{l c c}
\hline
Region & Percent change for lower (higher) $p_{\text{T, assoc}}^{\text{h,}\Lambda}$ & Lower (higher) $p_{\text{T, assoc}}^{\text{h,}\Lambda}$ significance \\
\hline
h--$\Lambda$ near-side &  $47.9 \pm 16.8$ ($46.6 \pm 14.6$) & $ 2.9\sigma (3.2\sigma) $\\ 
h--$\Lambda$ away-side &  $71.0 \pm 22.5$ ($46.2 \pm 17.9$) & $3.2\sigma (2.6\sigma)$ \\
h--h near-side &  $ 0.4 \pm 7.5$ ($-3.9 \pm 4.3$) & $0.1\sigma (-0.9\sigma)$ \\
h--h away-side &  $11.7 \pm 12.3$ ($1.0 \pm 7.0$) & $0.9\sigma (0.1\sigma)$ \\
\hline
\end{tabular}
\label{tab:percent_increase}
\end{table}

The per-trigger near- and away-side yields predicted by DPMJET are mostly consistent with data in the dihadron case. This can be seen in the model/data ratio, with both the near- and away-side ratios remaining close to unity across the entire multiplicity range. The h--$\Lambda$ yields, however, are not well described by the model. Both the near- and away-side h--$\Lambda$ yields predicted by DPMJET are lower than data by a factor of 2--2.5 depending on multiplicity in both momentum ranges, and there is no significant increase in these yields as a function of multiplicity. 

\begin{figure}[h!]
\centering
\includegraphics[width=\textwidth]{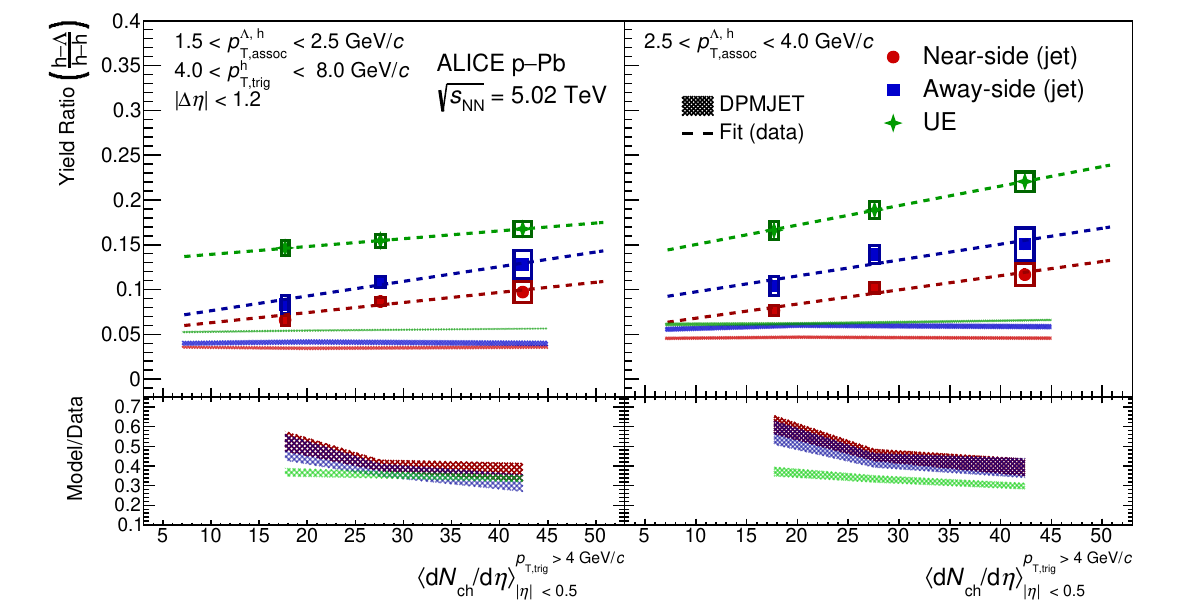}
\caption{The per-trigger pair-wise yield ratios $R_{i}^{\Lambda/\text{h}} \equiv Y_{i}^{\text{h--}\Lambda}$/$Y_{i}^{\text{h--h}}$ ($i$ = near-side, away-side, UE) as a function of multiplicity in the lower (left) and higher (right) associated momentum ranges. The statistical (systematic) uncertainties are shown as vertical lines (boxes). Linear fits to the data are given as dashed lines. The ratios predicted by DPMJET are presented as shaded bands, with the width of the band representing the uncertainty of the model. The ratios of the model to the data are also shown as shaded bands in the bottom panels, where the width of the band represents the uncertainty of the ratio.}
\label{fig:lambda_hadron_ratio}
\end{figure}

To better understand the differences between $\Lambda$ and charged hadron production both in and out of jets, the per-trigger yield ratios $R_{i}^{\Lambda/\text{h}} \equiv Y_{i}^{\text{h--}\Lambda}$/$Y_{i}^{\text{h--h}}$ ($i$ = near-side, away-side, UE) are measured as a function of multiplicity in both associated momentum ranges. These ratios serve as a proxy for the $\Lambda/\pi$ ratio in each region, and are shown in Fig.~\ref{fig:lambda_hadron_ratio}. Linear fits to the data are shown as dashed lines, with slopes and corresponding uncertainties reported in Table~\ref{tab:lambda_hadron_slopes}. The same ratios predicted by DPMJET are shown, with their ratios to the data presented in the bottom panels. 

A remarkable feature of these results is the clear separation between the yield ratios in each region across the entire multiplicity range in both momentum ranges, with the UE ratio being the largest, followed by the away-side ratio, and finally the near-side ratio. This indicates that most of the relative $\Lambda$ production is occurring in the UE, which is consistent with the idea that strange-quark production is maximal in this soft regime. This is further supported by the fact that the away-side ratio is larger than the near-side ratio, as $\Lambda$ production on the away-side is likely due to both the fragmentation of the away-side jet coupled with the possible production of strange quarks due to soft inelastic scattering against UE particles. Interestingly, DPMJET is able to produce this ordering in the ratios, but neither the magnitude nor the centrality dependence of the ratios are well described. In the lower-\pt range, the near- and away-side ratios predicted by DPMJET are lower than data by around a factor of 2 at the lowest multiplicities, which increases to a factor of 3 at the highest multiplicities. The higher-\pt interval shows a similar trend in the model/data ratios for both the near- and away-sides, but the magnitude of the deviation is smaller ($\approx 40$\% at the lowest multiplicities, $\approx 60$\% at the highest). The UE ratios predicted by DPMJET are also lower than data by around a factor of 3 across the entire multiplicity range in both momentum ranges, with the deviations being slightly larger at higher multiplicities.

\begin{table}
\centering
\caption{The slopes obtained from the linear fits to the per-trigger pair-wise (h--$\Lambda$)/(h--h) yield ratios as a function of multiplicity in both associated momentum ranges. The fits are made using the statistical and systematic uncertainties summed in quadrature. All fits are such that $\chi^{2}/\text{ndf} < 1$.}
\begin{tabular}{l c c}
\hline
Region & Lower-$p_{\text{T, assoc}}^{\text{h,}\Lambda}$ slope ($\times10^{-3}$) & Higher-$p_{\text{T, assoc}}^{\text{h,}\Lambda}$  slope ($\times10^{-3}$) \\
\hline
Near-side & $1.1 \pm 0.4$ & $1.6 \pm 0.4$ \\
Away-side & $1.6 \pm 0.6$ & $1.8 \pm 0.7$ \\
UE & $0.9 \pm 0.1$ & $2.2 \pm 0.2$ \\
\hline
\end{tabular}
\label{tab:lambda_hadron_slopes}
\end{table}

The near- and away-side slopes reported in Table~\ref{tab:lambda_hadron_slopes} are all more than $2\sigma$ greater than zero, hinting that there is an enhancement of relative $\Lambda$ production in jets as a function of multiplicity. This result is consistent with previous measurements of the $\phi(1020)$ meson in jets~\cite{ALICE:2024aid}, where a similar enhancement of the $\phi$/h ratio is observed in the near- and away-side regions. This further supports that the production of strange quarks is enhanced in jets in central \pPb collisions when compared to peripheral collisions in the momentum ranges considered. The UE slopes are also not compatible with zero, and the larger values of the UE ratios overall still suggest that a significant portion of the observed enhancement in the $\Lambda$/$\pi$ ratio with respect to multiplicity in Refs.~\cite{ALICEppEnhancement, ALICEpPbEnhancement} is due to soft production from the UE. Interestingly, the slopes for the UE region in the higher-\pt range are larger than in the lower-\pt range by over a factor of 2. This difference between the lower- and higher-\pt intervals could be due to an enhanced baryon production in the higher-\pt range, similar to that observed in Ref.~\cite{Lambda2}. The slopes calculated using the ratios obtained from DPMJET are all approximately zero, and are thus not shown in the table.

\subsection{Near- and away-side peak widths}

\begin{figure}[h!]
\centering
\includegraphics[width=\textwidth]{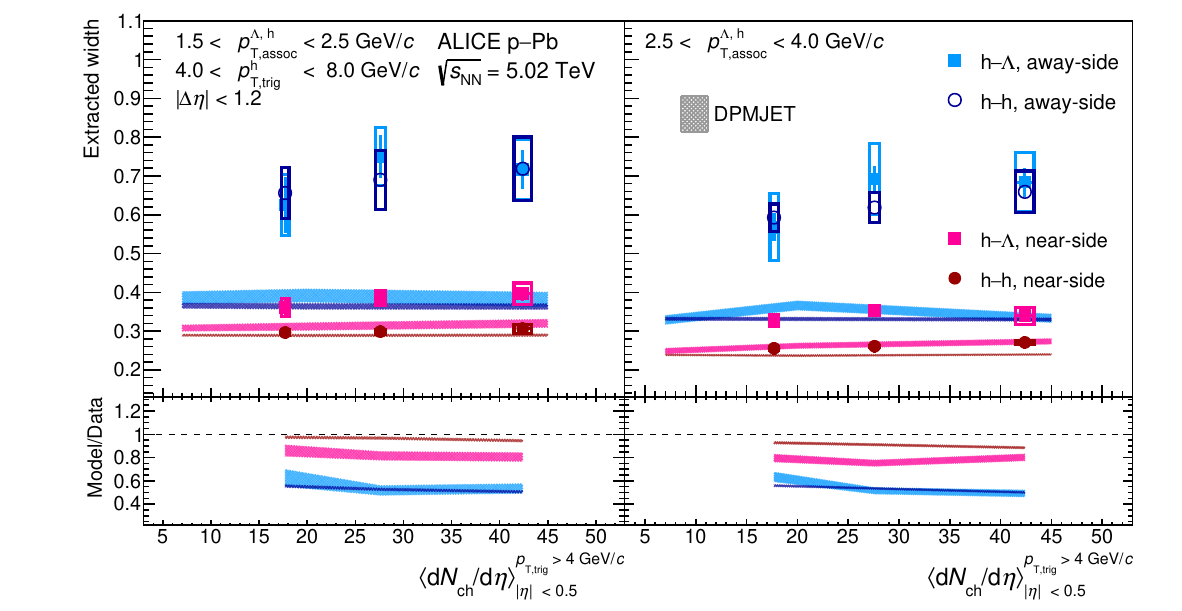}
\caption{The h--$\Lambda$ and h--h near- and away-side peak widths shown as a function of multiplicity for both associated momentum ranges. The statistical (systematic) uncertainties are shown as vertical lines (boxes). The widths predicted by DPMJET are presented as shaded bands, with the width of the band representing the uncertainty of the model. The ratios of the model to the data are also shown as shaded bands in the bottom panels, where the width of the band represents the uncertainty of the ratio. A dashed line is drawn at unity for reference.}
\label{fig:jet_widths}
\end{figure}

To gain more insight into the underlying mechanisms responsible for strangeness production in jets, the widths of the near- and away-side peaks are extracted from the h--$\Lambda$ and h--h $\Delta\varphi$ distributions using Eqs.~\ref{eq:fullfit} and~\ref{eq:width}. Plots of these widths as a function of multiplicity for both associated momentum ranges are shown in Fig.~\ref{fig:jet_widths}, along with the same widths predicted by DPMJET. A ratio of the model to the data is also presented in the bottom panels of the figure.

Expectedly, the near-side widths exhibit a slight decrease (around 10\%) from the lower-momentum range to the higher for both the h--$\Lambda$ and h--h cases, indicating that higher momentum associated particles are more collimated along the jet axis. An interesting result comes from comparing the h--$\Lambda$ and h--h away-side peak widths in data, which are found to be the same within systematic uncertainties across all multiplicity and momentum ranges, although the uncertainties are very large. This contrasts with the h--$\Lambda$ near-side widths, which are around $2\sigma$ larger than the h--h widths across the entire multiplicity range for both momentum ranges. For the h--h near-side widths, DPMJET describes the data well across both momentum ranges, with a $<5 (10)$\%  deviation from data seen in the lower- (higher-) momentum range. DPMJET also predicts the h--$\Lambda$ near-side width to be larger than the h--h width, though the values of the h--$\Lambda$ widths are much lower than they are in data. One explanation for these differences between the h--h and h--\lmb near-side widths, which are both observed in data and expected by DPMJET, could be due to the presence of gluon jets, which are generally wider than quark jets~\cite{GluonJet1}. As gluon jets exhibit an increased production of $\Lambda$ baryons~\cite{GluonJet2}, selecting on $\Lambda$ production---as is the case with the h--\lmb correlations---would bias the sample towards gluon jets, resulting in a larger near-side width when compared to the dihadron case.

DPMJET also under-predicts both the h--$\Lambda$ and h--h away-side widths by around 40\% across both momentum ranges. One explanation for this discrepancy could be that the away-side jet in data is subject to more collisional broadening than in the model due to the soft scattering against the UE particles, which are under-predicted by DPMJET. If this were the case, the discrepancy between data and DPMJET should increase with increasing multiplicity. However, as the away-side widths are consistent with flat behavior with respect to multiplicity due to the large systematic uncertainties, this possible explanation cannot be confirmed.

\subsection{Comparison with the $\phi(1020)$}

\begin{figure}[h!]
\centering
\includegraphics[width=\textwidth]{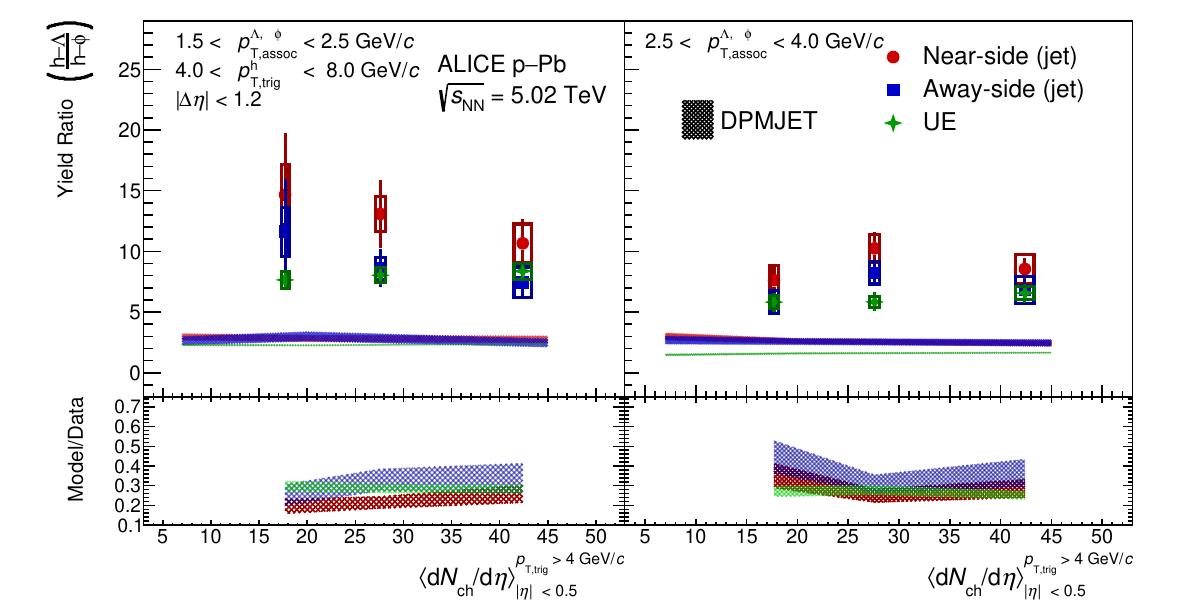}
\caption{The per-trigger pair-wise yield ratios $R_{i}^{\Lambda/\phi} \equiv Y_{i}^{\text{h--}\Lambda}$/$Y_{i}^{\text{h--}\phi}$ ($i$ = near-side, away-side, UE) as a function of multiplicity in the lower (left) and higher (right) associated momentum ranges. The statistical (systematic) uncertainties are shown as vertical lines (boxes). The ratios predicted by DPMJET are shown as shaded bands, with the width of the band representing the uncertainty of the model. The ratios of the model to the data are also given as shaded bands in the bottom panels, where the width of the band represents the uncertainty of the ratio.}
\label{fig:lambda_phi_ratio}
\end{figure}

Comparing the jet-like and UE production of $\Lambda$ baryons with other strange hadrons can provide further insight into the mechanisms responsible for strangeness production in \pPb collisions. The $\phi(1020)$ meson is of particular interest because even though it has a net strangeness $|S| = 0$, it has been observed to exhibit a similar enhancement in production as a function of multiplicity as other hadrons with non-zero strangeness~\cite{ALICE:2022wpn}. Due to their similar masses ($\Delta M < 100$ \MeVmass), the $\phi(1020)$ is an excellent candidate to compare directly with the $\Lambda$ in order to better understand the differences between open ($|S| \neq 0$) and hidden ($|S| = 0$) strange hadron production. Furthermore, the baryon-to-meson ratio is mostly independent of collision centrality in \pPb collisions in the momentum intervals considered in this analysis~\cite{ALICEpPbEnhancement}, indicating that any observed trends versus multiplicity are likely due to the strangeness content of the hadrons rather than the differing hadronization mechanisms. Using previously published results on $\phi(1020)$ production in and out of jets in \pPb collisions at $\sqrt{s_{\text{NN}}} = 5.02$ \TeV ~\cite{ALICE:2024aid}, the per-trigger pair-wise yield ratios $R_{i}^{\Lambda/\phi} \equiv Y_{i}^{\text{h--}\Lambda}/Y_{i}^{\text{h--}\phi}$ ($i$ = near-side, away-side, UE)  are obtained as a function of multiplicity, and are shown in Fig.~\ref{fig:lambda_phi_ratio} for both the lower and higher associated \pt ranges. The h--$\phi$ yields and their corresponding systematic and statistical uncertainties are taken directly from the published results. The uncertainties for $Y_{i}^{\text{h--}\Lambda}$ and $Y_{i}^{\text{h--}\phi}$ are treated as uncorrelated when computing the uncertainties in the ratios. The UE yields are extracted using the techniques described in Section~\ref{sec:uefit}. The same ratios predicted by DPMJET are also presented, with their ratios to the data shown in the bottom panels.

Interestingly, the $\Lambda/\phi$ near-side ratios appear to be systematically higher than the ratios in the UE, with the significance of the difference varying from $0.9\sigma$ where the ratios are the closest, up to $2.5\sigma$ when they are most separated. This hints at a relative enhancement (suppression) of $\Lambda$ ($\phi$) production along the jet axis. As strangeness in QCD can only be produced in the form of $\text{s}\overline{\text{s}}$ pairs, one possible explanation of this effect is that when these pairs are produced from an initial hard scattering of the form $\text{q}\overline{\text{q}} \rightarrow \text{s}\overline{\text{s}}$ (i.e., the lowest order pQCD diagram for quark-based strangeness production from Ref.~\cite{QGPStrangeness}), the $\text{s}$ and $\overline{\text{s}}$ will likely never hadronize into the same $\phi$ due to their separation in phase space. The ratios predicted by DPMJET provide further evidence for this explanation, as the model also shows that the near-side ratios are enhanced relative to the UE ratios across the entire multiplicity range. However, DPMJET does not predict the differences between the near- and away-side $\Lambda$/$\phi$ ratios observed in data, in which the near-side ratios are systematically higher than the away-side ones. This could be explained by an increase of strange-quark production in the away-side due to soft scattering against the UE particles that are under-predicted in the model. These strange quarks would occupy a similar phase-space as the strange quark produced from the initial fragmentation, increasing the chances for an $\text{s}\overline{\text{s}}$ pair to hadronize into a $\phi$. While the values of the near- and away-side ratios in the lower associated momentum range decrease with increasing multiplicity, the uncertainties are very large. As a result, no definitive conclusions can be drawn about the multiplicity dependence of the $\Lambda/\phi$ ratios in the jet and in the UE.

\section{Conclusion}
\label{sec:conclusion}
This article presents the first results of $\Lambda$ production in p--Pb collisions at \snn $= 5.02$ \TeV in the jet and in the underlying event extracted through angular correlation measurements, using a high-\pt charged hadron as a proxy for the jet axis. Both the per-trigger yields in each kinematic region and the widths of the near- and away-side peaks are extracted from the azimuthal correlation functions and studied as a function of associated-particle \pt and event multiplicity. 
The associated \lmb yields show an increase with increasing multiplicity with a significance of about $3\sigma$ for both the near and away sides, whereas the charged hadron associated yields in the near and away sides do not exhibit a statistically significant multiplicity dependence.

The per-trigger pair-wise yield ratios $R_{i}^{\Lambda/\text{h}}$ and $R_{i}^{\Lambda/\phi}$ ($i$ = near-side, away-side, UE) are also studied as a function of associated-particle \pt and multiplicity. The $\Lambda/\text{h}$ ratios exhibit a clear ordering in each region for the entire multiplicity range in both \pt ranges, with the UE ratios being larger than the away-side ratios, which are larger than the near-side ratios. The $\Lambda/\text{h}$ ratios in each region also increase with multiplicity, with slopes that are greater than zero by over $2\sigma$ for both momentum ranges. The $\Lambda/\phi$ ratios in the near-side jet region are measured to be systematically higher than both the away-side and UE ratios, hinting at a suppression of $\phi$ mesons with respect to the $\Lambda$ along the jet axis relative to those produced out of jet. Furthermore, the $\Lambda/\phi$ ratios show no significant dependence on multiplicity in both associated-particle momentum ranges. The $\Lambda/\text{h}$ results suggest that, while strange quark production occurs mostly in the UE, the observed enhancement of strangeness in high-multiplicity \pPb and \PbPb collisions when compared to low-multiplicity pp collisions is at least partially driven by jet-like processes. Additionally, the $\Lambda/\phi$ results provide some insight into the phase-space separation of $\text{s}\bar{\text{s}}$ pairs in the jet and UE regions, and suggest that $\text{s}(\bar{\text{s}})$ quarks produced along the jet axis may be less likely to hadronize with another $\bar{\text{s}} (\text{s})$ quark to form a $\phi$ meson when compared to those produced out of jet.

The h--$\Lambda$ and h--h near-side peak widths reveal a large dependence on \pt, with the peaks becoming more collimated as momentum increases. The widths of the away-side jets are found to be independent of both \pt and multiplicity, with the larger systematic uncertainties preventing exclusion of flat behavior. Comparing the widths of h--$\Lambda$ and h--h correlations reveals that the h--$\Lambda$ near-side widths are significantly ($>2\sigma$) larger than the h--h near-side widths, hinting at a bias towards gluon jets in the h--$\Lambda$ sample. However, the away-side widths are found to be consistent within uncertainties between the two samples.

All measured observables are compared with predictions from the DPMJET model. The predicted near- and away-side yields are found to be in relatively good agreement with data in the dihadron case, but the h--$\Lambda$ yield predictions deviate from data by a large ($>40$\%) margin. DPMJET also fails to predict any of the observed multiplicity dependence for the h--$\Lambda$ near- and away-side yields. However, the model is able to closely predict the near-side widths of the dihadron distributions across all multiplicity and momentum ranges, although it under-predicts both the h--$\Lambda$ near-side widths and the away-side widths for both ($\Lambda$, h) cases. The model also predicts a difference between the h--$\Lambda$ and h--h near-side widths, which is observed in data as well. The per-trigger $\Lambda/\text{h}$ and $\Lambda/\phi$ yield ratios are consistently under-predicted by DPMJET, and exhibit no multiplicity dependence. Even still, DPMJET manages to predict the ordering of the $\Lambda/\text{h}$ ratios in each region (UE $>$ away-side jet $>$ near-side jet) and the enhancement of the near- and away-side $\Lambda/\phi$ ratios when compared to the UE region. 

Using the data collected during LHC Run 3 (2022--2025), the measurements presented in this paper can be extended by utilizing similar techniques to study the production of multi-strange hadrons (e.g., $\Xi^-$ and $\Omega^-$) in p--Pb collisions. Such measurements would provide further insight into the production mechanisms of strange hadrons in small collision systems, and would help to constrain the models used to describe these systems.


\newenvironment{acknowledgement}{\relax}{\relax}
\begin{acknowledgement}
\section*{Acknowledgements}

The ALICE Collaboration would like to thank all its engineers and technicians for their invaluable contributions to the construction of the experiment and the CERN accelerator teams for the outstanding performance of the LHC complex.
The ALICE Collaboration gratefully acknowledges the resources and support provided by all Grid centres and the Worldwide LHC Computing Grid (WLCG) collaboration.
The ALICE Collaboration acknowledges the following funding agencies for their support in building and running the ALICE detector:
A. I. Alikhanyan National Science Laboratory (Yerevan Physics Institute) Foundation (ANSL), State Committee of Science and World Federation of Scientists (WFS), Armenia;
Austrian Academy of Sciences, Austrian Science Fund (FWF): [M 2467-N36] and Nationalstiftung f\"{u}r Forschung, Technologie und Entwicklung, Austria;
Ministry of Communications and High Technologies, National Nuclear Research Center, Azerbaijan;
Conselho Nacional de Desenvolvimento Cient\'{\i}fico e Tecnol\'{o}gico (CNPq), Financiadora de Estudos e Projetos (Finep), Funda\c{c}\~{a}o de Amparo \`{a} Pesquisa do Estado de S\~{a}o Paulo (FAPESP) and Universidade Federal do Rio Grande do Sul (UFRGS), Brazil;
Bulgarian Ministry of Education and Science, within the National Roadmap for Research Infrastructures 2020-2027 (object CERN), Bulgaria;
Ministry of Education of China (MOEC) , Ministry of Science \& Technology of China (MSTC) and National Natural Science Foundation of China (NSFC), China;
Ministry of Science and Education and Croatian Science Foundation, Croatia;
Centro de Aplicaciones Tecnol\'{o}gicas y Desarrollo Nuclear (CEADEN), Cubaenerg\'{\i}a, Cuba;
Ministry of Education, Youth and Sports of the Czech Republic, Czech Republic;
The Danish Council for Independent Research | Natural Sciences, the VILLUM FONDEN and Danish National Research Foundation (DNRF), Denmark;
Helsinki Institute of Physics (HIP), Finland;
Commissariat \`{a} l'Energie Atomique (CEA) and Institut National de Physique Nucl\'{e}aire et de Physique des Particules (IN2P3) and Centre National de la Recherche Scientifique (CNRS), France;
Bundesministerium f\"{u}r Bildung und Forschung (BMBF) and GSI Helmholtzzentrum f\"{u}r Schwerionenforschung GmbH, Germany;
General Secretariat for Research and Technology, Ministry of Education, Research and Religions, Greece;
National Research, Development and Innovation Office, Hungary;
Department of Atomic Energy Government of India (DAE), Department of Science and Technology, Government of India (DST), University Grants Commission, Government of India (UGC) and Council of Scientific and Industrial Research (CSIR), India;
National Research and Innovation Agency - BRIN, Indonesia;
Istituto Nazionale di Fisica Nucleare (INFN), Italy;
Japanese Ministry of Education, Culture, Sports, Science and Technology (MEXT) and Japan Society for the Promotion of Science (JSPS) KAKENHI, Japan;
Consejo Nacional de Ciencia (CONACYT) y Tecnolog\'{i}a, through Fondo de Cooperaci\'{o}n Internacional en Ciencia y Tecnolog\'{i}a (FONCICYT) and Direcci\'{o}n General de Asuntos del Personal Academico (DGAPA), Mexico;
Nederlandse Organisatie voor Wetenschappelijk Onderzoek (NWO), Netherlands;
The Research Council of Norway, Norway;
Pontificia Universidad Cat\'{o}lica del Per\'{u}, Peru;
Ministry of Science and Higher Education, National Science Centre and WUT ID-UB, Poland;
Korea Institute of Science and Technology Information and National Research Foundation of Korea (NRF), Republic of Korea;
Ministry of Education and Scientific Research, Institute of Atomic Physics, Ministry of Research and Innovation and Institute of Atomic Physics and Universitatea Nationala de Stiinta si Tehnologie Politehnica Bucuresti, Romania;
Ministry of Education, Science, Research and Sport of the Slovak Republic, Slovakia;
National Research Foundation of South Africa, South Africa;
Swedish Research Council (VR) and Knut \& Alice Wallenberg Foundation (KAW), Sweden;
European Organization for Nuclear Research, Switzerland;
Suranaree University of Technology (SUT), National Science and Technology Development Agency (NSTDA) and National Science, Research and Innovation Fund (NSRF via PMU-B B05F650021), Thailand;
Turkish Energy, Nuclear and Mineral Research Agency (TENMAK), Turkey;
National Academy of  Sciences of Ukraine, Ukraine;
Science and Technology Facilities Council (STFC), United Kingdom;
National Science Foundation of the United States of America (NSF) and United States Department of Energy, Office of Nuclear Physics (DOE NP), United States of America.
In addition, individual groups or members have received support from:
Czech Science Foundation (grant no. 23-07499S), Czech Republic;
European Research Council (grant no. 950692), European Union;
ICSC - Centro Nazionale di Ricerca in High Performance Computing, Big Data and Quantum Computing, European Union - NextGenerationEU;
Academy of Finland (Center of Excellence in Quark Matter) (grant nos. 346327, 346328), Finland.

\end{acknowledgement}

\bibliographystyle{utphys}   
\bibliography{bibliography}

\newpage
\appendix

%
%

\section{The ALICE Collaboration}
\label{app:collab}
\begin{flushleft} 
\small

S.~Acharya\,\orcidlink{0000-0002-9213-5329}\,$^{\rm 127}$, 
D.~Adamov\'{a}\,\orcidlink{0000-0002-0504-7428}\,$^{\rm 86}$, 
A.~Agarwal$^{\rm 135}$, 
G.~Aglieri Rinella\,\orcidlink{0000-0002-9611-3696}\,$^{\rm 32}$, 
L.~Aglietta\,\orcidlink{0009-0003-0763-6802}\,$^{\rm 24}$, 
M.~Agnello\,\orcidlink{0000-0002-0760-5075}\,$^{\rm 29}$, 
N.~Agrawal\,\orcidlink{0000-0003-0348-9836}\,$^{\rm 25}$, 
Z.~Ahammed\,\orcidlink{0000-0001-5241-7412}\,$^{\rm 135}$, 
S.~Ahmad\,\orcidlink{0000-0003-0497-5705}\,$^{\rm 15}$, 
S.U.~Ahn\,\orcidlink{0000-0001-8847-489X}\,$^{\rm 71}$
% ...

\section*{Affiliation Notes}

$^{\rm I}$ Deceased\\
$^{\rm II}$ Also at: Max-Planck-Institut fur Physik, Munich, Germany\\
$^{\rm III}$ Also at: Italian National Agency for New Technologies, Energy and Sustainable Economic Development (ENEA), Bologna, Italy\\
$^{\rm IV}$ Also at: Dipartimento DET del Politecnico di Torino, Turin, Italy\\
$^{\rm V}$ Also at: Yildiz Technical University, Istanbul, T\"{u}rkiye\\
$^{\rm VI}$ Also at: Department of Applied Physics, Aligarh Muslim University, Aligarh, India\\
$^{\rm VII}$ Also at: Institute of Theoretical Physics, University of Wroclaw, Poland\\
$^{\rm VIII}$ Also at: An institution covered by a cooperation agreement with CERN\\

\section*{Collaboration Institutes}

$^{1}$ A.I. Alikhanyan National Science Laboratory (Yerevan Physics Institute) Foundation, Yerevan, Armenia\\
$^{2}$ AGH University of Krakow, Cracow, Poland\\
$^{3}$ Bogolyubov Institute for Theoretical Physics, National Academy of Sciences of Ukraine, Kiev, Ukraine\\
$^{4}$ Bose Institute, Department of Physics  and Centre for Astroparticle Physics and Space Science (CAPSS), Kolkata, India\\
$^{5}$ California Polytechnic State University, San Luis Obispo, California, United States\\
$^{6}$ Central China Normal University, Wuhan, China\\
$^{7}$ Centro de Aplicaciones Tecnol\'{o}gicas y Desarrollo Nuclear (CEADEN), Havana, Cuba\\
$^{8}$ Centro de Investigaci\'{o}n y de Estudios Avanzados (CINVESTAV), Mexico City and M\'{e}rida, Mexico\\
$^{9}$ Chicago State University, Chicago, Illinois, United States\\
$^{10}$ China Institute of Atomic Energy, Beijing, China\\
$^{11}$ China University of Geosciences, Wuhan, China\\
$^{12}$ Chungbuk National University, Cheongju, Republic of Korea\\
$^{13}$ Comenius University Bratislava, Faculty of Mathematics, Physics and Informatics, Bratislava, Slovak Republic\\
$^{14}$ Creighton University, Omaha, Nebraska, United States\\
$^{15}$ Department of Physics, Aligarh Muslim University, Aligarh, India\\
$^{16}$ Department of Physics, Pusan National University, Pusan, Republic of Korea\\
$^{17}$ Department of Physics, Sejong University, Seoul, Republic of Korea\\
$^{18}$ Department of Physics, University of California, Berkeley, California, United States\\
$^{19}$ Department of Physics, University of Oslo, Oslo, Norway\\
$^{20}$ Department of Physics and Technology, University of Bergen, Bergen, Norway\\
$^{21}$ Dipartimento di Fisica, Universit\`{a} di Pavia, Pavia, Italy\\
$^{22}$ Dipartimento di Fisica dell'Universit\`{a} and Sezione INFN, Cagliari, Italy\\
$^{23}$ Dipartimento di Fisica dell'Universit\`{a} and Sezione INFN, Trieste, Italy\\
$^{24}$ Dipartimento di Fisica dell'Universit\`{a} and Sezione INFN, Turin, Italy\\
$^{25}$ Dipartimento di Fisica e Astronomia dell'Universit\`{a} and Sezione INFN, Bologna, Italy\\
$^{26}$ Dipartimento di Fisica e Astronomia dell'Universit\`{a} and Sezione INFN, Catania, Italy\\
$^{27}$ Dipartimento di Fisica e Astronomia dell'Universit\`{a} and Sezione INFN, Padova, Italy\\
$^{28}$ Dipartimento di Fisica `E.R.~Caianiello' dell'Universit\`{a} and Gruppo Collegato INFN, Salerno, Italy\\
$^{29}$ Dipartimento DISAT del Politecnico and Sezione INFN, Turin, Italy\\
$^{30}$ Dipartimento di Scienze MIFT, Universit\`{a} di Messina, Messina, Italy\\
$^{31}$ Dipartimento Interateneo di Fisica `M.~Merlin' and Sezione INFN, Bari, Italy\\
$^{32}$ European Organization for Nuclear Research (CERN), Geneva, Switzerland\\
$^{33}$ Faculty of Electrical Engineering, Mechanical Engineering and Naval Architecture, University of Split, Split, Croatia\\
$^{34}$ Faculty of Engineering and Science, Western Norway University of Applied Sciences, Bergen, Norway\\
$^{35}$ Faculty of Nuclear Sciences and Physical Engineering, Czech Technical University in Prague, Prague, Czech Republic\\
$^{36}$ Faculty of Physics, Sofia University, Sofia, Bulgaria\\
$^{37}$ Faculty of Science, P.J.~\v{S}af\'{a}rik University, Ko\v{s}ice, Slovak Republic\\
$^{38}$ Frankfurt Institute for Advanced Studies, Johann Wolfgang Goethe-Universit\"{a}t Frankfurt, Frankfurt, Germany\\
$^{39}$ Fudan University, Shanghai, China\\
$^{40}$ Gangneung-Wonju National University, Gangneung, Republic of Korea\\
$^{41}$ Gauhati University, Department of Physics, Guwahati, India\\
$^{42}$ Helmholtz-Institut f\"{u}r Strahlen- und Kernphysik, Rheinische Friedrich-Wilhelms-Universit\"{a}t Bonn, Bonn, Germany\\
$^{43}$ Helsinki Institute of Physics (HIP), Helsinki, Finland\\
$^{44}$ High Energy Physics Group,  Universidad Aut\'{o}noma de Puebla, Puebla, Mexico\\
$^{45}$ Horia Hulubei National Institute of Physics and Nuclear Engineering, Bucharest, Romania\\
$^{46}$ HUN-REN Wigner Research Centre for Physics, Budapest, Hungary\\
$^{47}$ Indian Institute of Technology Bombay (IIT), Mumbai, India\\
$^{48}$ Indian Institute of Technology Indore, Indore, India\\
$^{49}$ INFN, Laboratori Nazionali di Frascati, Frascati, Italy\\
$^{50}$ INFN, Sezione di Bari, Bari, Italy\\
$^{51}$ INFN, Sezione di Bologna, Bologna, Italy\\
$^{52}$ INFN, Sezione di Cagliari, Cagliari, Italy\\
$^{53}$ INFN, Sezione di Catania, Catania, Italy\\
$^{54}$ INFN, Sezione di Padova, Padova, Italy\\
$^{55}$ INFN, Sezione di Pavia, Pavia, Italy\\
$^{56}$ INFN, Sezione di Torino, Turin, Italy\\
$^{57}$ INFN, Sezione di Trieste, Trieste, Italy\\
$^{58}$ Inha University, Incheon, Republic of Korea\\
$^{59}$ Institute for Gravitational and Subatomic Physics (GRASP), Utrecht University/Nikhef, Utrecht, Netherlands\\
$^{60}$ Institute of Experimental Physics, Slovak Academy of Sciences, Ko\v{s}ice, Slovak Republic\\
$^{61}$ Institute of Physics, Homi Bhabha National Institute, Bhubaneswar, India\\
$^{62}$ Institute of Physics of the Czech Academy of Sciences, Prague, Czech Republic\\
$^{63}$ Institute of Space Science (ISS), Bucharest, Romania\\
$^{64}$ Institut f\"{u}r Kernphysik, Johann Wolfgang Goethe-Universit\"{a}t Frankfurt, Frankfurt, Germany\\
$^{65}$ Instituto de Ciencias Nucleares, Universidad Nacional Aut\'{o}noma de M\'{e}xico, Mexico City, Mexico\\
$^{66}$ Instituto de F\'{i}sica, Universidade Federal do Rio Grande do Sul (UFRGS), Porto Alegre, Brazil\\
$^{67}$ Instituto de F\'{\i}sica, Universidad Nacional Aut\'{o}noma de M\'{e}xico, Mexico City, Mexico\\
$^{68}$ iThemba LABS, National Research Foundation, Somerset West, South Africa\\
$^{69}$ Jeonbuk National University, Jeonju, Republic of Korea\\
$^{70}$ Johann-Wolfgang-Goethe Universit\"{a}t Frankfurt Institut f\"{u}r Informatik, Fachbereich Informatik und Mathematik, Frankfurt, Germany\\
$^{71}$ Korea Institute of Science and Technology Information, Daejeon, Republic of Korea\\
$^{72}$ KTO Karatay University, Konya, Turkey\\
$^{73}$ Laboratoire de Physique Subatomique et de Cosmologie, Universit\'{e} Grenoble-Alpes, CNRS-IN2P3, Grenoble, France\\
$^{74}$ Lawrence Berkeley National Laboratory, Berkeley, California, United States\\
$^{75}$ Lund University Department of Physics, Division of Particle Physics, Lund, Sweden\\
$^{76}$ Nagasaki Institute of Applied Science, Nagasaki, Japan\\
$^{77}$ Nara Women{'}s University (NWU), Nara, Japan\\
$^{78}$ National and Kapodistrian University of Athens, School of Science, Department of Physics , Athens, Greece\\
$^{79}$ National Centre for Nuclear Research, Warsaw, Poland\\
$^{80}$ National Institute of Science Education and Research, Homi Bhabha National Institute, Jatni, India\\
$^{81}$ National Nuclear Research Center, Baku, Azerbaijan\\
$^{82}$ National Research and Innovation Agency - BRIN, Jakarta, Indonesia\\
$^{83}$ Niels Bohr Institute, University of Copenhagen, Copenhagen, Denmark\\
$^{84}$ Nikhef, National institute for subatomic physics, Amsterdam, Netherlands\\
$^{85}$ Nuclear Physics Group, STFC Daresbury Laboratory, Daresbury, United Kingdom\\
$^{86}$ Nuclear Physics Institute of the Czech Academy of Sciences, Husinec-\v{R}e\v{z}, Czech Republic\\
$^{87}$ Oak Ridge National Laboratory, Oak Ridge, Tennessee, United States\\
$^{88}$ Ohio State University, Columbus, Ohio, United States\\
$^{89}$ Physics department, Faculty of science, University of Zagreb, Zagreb, Croatia\\
$^{90}$ Physics Department, Panjab University, Chandigarh, India\\
$^{91}$ Physics Department, University of Jammu, Jammu, India\\
$^{92}$ Physics Program and International Institute for Sustainability with Knotted Chiral Meta Matter (SKCM2), Hiroshima University, Hiroshima, Japan\\
$^{93}$ Physikalisches Institut, Eberhard-Karls-Universit\"{a}t T\"{u}bingen, T\"{u}bingen, Germany\\
$^{94}$ Physikalisches Institut, Ruprecht-Karls-Universit\"{a}t Heidelberg, Heidelberg, Germany\\
$^{95}$ Physik Department, Technische Universit\"{a}t M\"{u}nchen, Munich, Germany\\
$^{96}$ Politecnico di Bari and Sezione INFN, Bari, Italy\\
$^{97}$ Research Division and ExtreMe Matter Institute EMMI, GSI Helmholtzzentrum f\"ur Schwerionenforschung GmbH, Darmstadt, Germany\\
$^{98}$ Saga University, Saga, Japan\\
$^{99}$ Saha Institute of Nuclear Physics, Homi Bhabha National Institute, Kolkata, India\\
$^{100}$ School of Physics and Astronomy, University of Birmingham, Birmingham, United Kingdom\\
$^{101}$ Secci\'{o}n F\'{\i}sica, Departamento de Ciencias, Pontificia Universidad Cat\'{o}lica del Per\'{u}, Lima, Peru\\
$^{102}$ Stefan Meyer Institut f\"{u}r Subatomare Physik (SMI), Vienna, Austria\\
$^{103}$ SUBATECH, IMT Atlantique, Nantes Universit\'{e}, CNRS-IN2P3, Nantes, France\\
$^{104}$ Sungkyunkwan University, Suwon City, Republic of Korea\\
$^{105}$ Suranaree University of Technology, Nakhon Ratchasima, Thailand\\
$^{106}$ Technical University of Ko\v{s}ice, Ko\v{s}ice, Slovak Republic\\
$^{107}$ The Henryk Niewodniczanski Institute of Nuclear Physics, Polish Academy of Sciences, Cracow, Poland\\
$^{108}$ The University of Texas at Austin, Austin, Texas, United States\\
$^{109}$ Universidad Aut\'{o}noma de Sinaloa, Culiac\'{a}n, Mexico\\
$^{110}$ Universidade de S\~{a}o Paulo (USP), S\~{a}o Paulo, Brazil\\
$^{111}$ Universidade Estadual de Campinas (UNICAMP), Campinas, Brazil\\
$^{112}$ Universidade Federal do ABC, Santo Andre, Brazil\\
$^{113}$ Universitatea Nationala de Stiinta si Tehnologie Politehnica Bucuresti, Bucharest, Romania\\
$^{114}$ University of Cape Town, Cape Town, South Africa\\
$^{115}$ University of Derby, Derby, United Kingdom\\
$^{116}$ University of Houston, Houston, Texas, United States\\
$^{117}$ University of Jyv\"{a}skyl\"{a}, Jyv\"{a}skyl\"{a}, Finland\\
$^{118}$ University of Kansas, Lawrence, Kansas, United States\\
$^{119}$ University of Liverpool, Liverpool, United Kingdom\\
$^{120}$ University of Science and Technology of China, Hefei, China\\
$^{121}$ University of South-Eastern Norway, Kongsberg, Norway\\
$^{122}$ University of Tennessee, Knoxville, Tennessee, United States\\
$^{123}$ University of the Witwatersrand, Johannesburg, South Africa\\
$^{124}$ University of Tokyo, Tokyo, Japan\\
$^{125}$ University of Tsukuba, Tsukuba, Japan\\
$^{126}$ Universit\"{a}t M\"{u}nster, Institut f\"{u}r Kernphysik, M\"{u}nster, Germany\\
$^{127}$ Universit\'{e} Clermont Auvergne, CNRS/IN2P3, LPC, Clermont-Ferrand, France\\
$^{128}$ Universit\'{e} de Lyon, CNRS/IN2P3, Institut de Physique des 2 Infinis de Lyon, Lyon, France\\
$^{129}$ Universit\'{e} de Strasbourg, CNRS, IPHC UMR 7178, F-67000 Strasbourg, France, Strasbourg, France\\
$^{130}$ Universit\'{e} Paris-Saclay, Centre d'Etudes de Saclay (CEA), IRFU, D\'{e}partment de Physique Nucl\'{e}aire (DPhN), Saclay, France\\
$^{131}$ Universit\'{e}  Paris-Saclay, CNRS/IN2P3, IJCLab, Orsay, France\\
$^{132}$ Universit\`{a} degli Studi di Foggia, Foggia, Italy\\
$^{133}$ Universit\`{a} del Piemonte Orientale, Vercelli, Italy\\
$^{134}$ Universit\`{a} di Brescia, Brescia, Italy\\
$^{135}$ Variable Energy Cyclotron Centre, Homi Bhabha National Institute, Kolkata, India\\
$^{136}$ Warsaw University of Technology, Warsaw, Poland\\
$^{137}$ Wayne State University, Detroit, Michigan, United States\\
$^{138}$ Yale University, New Haven, Connecticut, United States\\
$^{139}$ Yonsei University, Seoul, Republic of Korea\\
$^{140}$  Zentrum  f\"{u}r Technologie und Transfer (ZTT), Worms, Germany\\
$^{141}$ Affiliated with an institute covered by a cooperation agreement with CERN\\
$^{142}$ Affiliated with an international laboratory covered by a cooperation agreement with CERN.\\

\end{flushleft} 

\end{document}